\documentclass[a4paper,fleqn,usenatbib]{mnras}
\usepackage{epsf,palatino,url,graphicx,wrapfig,array,setspace,amssymb,multirow,lscape,appendix,rotating,amsmath}

\usepackage{mathptmx}
\usepackage{txfonts}

\usepackage[T1]{fontenc}
\usepackage{ae,aecompl}

\bibpunct{(}{)}{;}{a}{}{,} 

\newcommand{\sxp}{SXP\,7.92}

\newcommand{\ergs}[1]{$\times 10^{#1}$ erg s$^{-1}$}
\newcommand{\unitF}{erg cm$^{-2}$ s$^{-1}$}
\newcommand{\cmsq}{cm$^{-2}$}
\newcommand{\unitL}{erg s$^{-1}$}
\newcommand{\nh}{N$_{\rm H}$}
\newcommand{\lx}{\hbox{L$_{\rm x}$}}

\newcommand{\SiIV}{\ion{Si}{IV}}

\newcommand{\HeII}{\ion{He}{II}}
\newcommand{\OII}{\ion{O}{II}}

\newcommand{\Halp}{H${\alpha}$}

\newcommand{\kms}{km~s$^{-1}$}

\newcommand{\cxo}{\emph{Chandra}}
\newcommand{\xmm}{\emph{XMM-Newton}}
\newcommand{\integral}{\emph{INTEGRAL}}
\newcommand{\rxte}{\emph{RXTE}}
\newcommand{\swift}{\emph{Swift}}

\newcommand{\aara}

\title[SXP7.92]{\sxp: A Recently Rediscovered Be/X-ray Binary in the Small Magellanic Cloud, Viewed Edge On}

\author[E.~S.~Bartlett et al.]{E.~S.~Bartlett,$^{1,2}$\thanks{E-mail: ebartlet@eso.org (ESB)}
M.~J.~Coe$^{3}$,
G.~L.~Israel$^{4}$,
J.~S. Clark$^{5}$,
P.~Esposito$^{6}$,
V.~D'Elia$^{4,7}$
\newauthor{and A.~Udalski$^{8}$}
\\
$^{1}$ESO - European Southern Observatory, Alonso de C\'{o}rdova 3107, Vitacura, Casilla 19001, Santiago de Chile, Chile\\
$^{2}$Astrophysics, Cosmology and Gravity Centre, Department of Astronomy, University of Cape Town, Rondebosch 7701, South Africa\\
$^{3}$School of Physics and Astronomy, University of Southampton, Highfield, Southampton, SO17 1BJ, United Kingdom\\
$^{4}$INAF -- Osservatorio Astronomico di Roma, via Frascati 33, I-00040 Monte Porzio Catone (Roma), Italy\\
$^{5}$Department of Physical Science, The Open University, Walton Hall, Milton Keynes, MK7 6AA, United Kingdom\\
$^{6}$Anton Pannekoek Institute for Astronomy, University of Amsterdam, Postbus 94249, NL-1090-GE Amsterdam, The Netherlands\\
$^{7}$ASI Science Data Center (ASDC), via del Politecnico snc, I-00133 Roma, Italy\\
$^{8}$Warsaw University Observatory, Aleje Ujazdowskie 4, 00-478 Warszawa, Poland
}

\date{Accepted 2017 January 05 . Received 2017 January 05 ; in original form 2016 July 26}

\pubyear{2017}

\begin{document}
\label{firstpage}
\pagerange{\pageref{firstpage}--\pageref{lastpage}}
\maketitle

\begin{abstract}

We present a detailed optical and X-ray study of the 2013 outburst of the Small Magellanic Cloud Be/X-ray binary \sxp, as well as an overview of the last 18 years of observations from OGLE, \rxte, \cxo\ and \xmm. We revise the position of this source to RA(J2000)=00:57:58.4, Dec(J2000)=-72:22:29.5 with a $1\sigma$ uncertainty of 1.5\arcsec, correcting the previously reported position by \citet{Coe2009} by more than 20 arcminutes. We identify and spectrally classify the correct counterpart as a B1Ve star. The optical spectrum is distinguished by an uncharacteristically deep narrow Balmer series, with the \Halp\ line in particular having a distinctive shell profile, i.e. a deep absorption core embedded in an emission line. We interpret this as evidence that we are viewing the system edge on and are seeing self obscuration of the circumstellar disc. We derive an optical period for the system of 40.0$\pm0.3$~days, which we interpret as the orbital period, and present several mechanisms to describe the X-ray/Optical behaviour in the recent outburst, in particular the ``flares'' and ``dips'' seen in the optical light curve, including a transient accretion disc and an elongated precessing disc.

\end{abstract}




\section{Introduction}\label{sect:intro}

High-Mass X-ray Binaries (HMXBs) are stellar systems in which a compact object, usually a neutron star (NS) with a strong magnetic field, orbits a massive star of spectral type OB. HMXBs broadly fall into two categories: Be/X-ray binaries (BeXRBs) in which the primary is a rapidly rotating early type star with luminosity class III-V, and supergiant X-ray Binaries (sgXRBs), in which the primary is a post main sequence star. Material is accreted onto the compact object, either via Roche-Lobe overflow in the case of the sgXRBs, or directly from the circumstellar disc that surrounds the Be star in the BeXRBs (and leads to their `e' designation). BeXRBs are currently the most numerous subclass of HMXB \citep{Liu2006}, though we note that, since the launch of \integral\ a new type of sgXRB has been identified; the Supergiant Fast X-ray Transients (SFXTs, \citealt{Negueruela2006}) whose numbers are increasing rapidly. With binarity increasingly being recognised as playing a major role in the evolution of massive stars \citep{Sana2012}, HMXBs represent a pivotal point in their evolutionary track: Once one star has gone supernova but whilst the other is still fusing hydrogen, either in its core or envelope. This, in turn, means that HMXBs could provide valuable insight into many astrophysical phenomena, including core collapse supernovae, gamma-ray bursts, magnetars and gravitational waves.

The `e' designation stands for emission: Be stars have, at some point in their lives, displayed one or more of the Balmer series in emission in their optical spectra. These lines originate from the equatorial outflow, or decretion disc, around the star. This disc is fed in part by the star's rapid rotation \citep{Townsend2004}, though the creation mechanism is still unclear (possibly linked to the transient nature of the emission lines). The accreted material is funnelled on the magnetic poles of the NSs. These NSs are rotating, leading these systems to be coherently variable X-ray sources, i.e. pulsators. The orbits of the NSs in these systems are highly eccentric, and are loosely correlated with the spin period \citep{Corbet1984,Corbet1986}. There are two types of X-ray outburst traditionally associated with BeXRBs, Type I outbursts, which occur around the time of periastron passage, last days-weeks and reach \lx $\sim10^{36-37}$ \unitL and Type II ``Giant'' outbursts which last longer (up to months) reach higher luminosities (\lx $>10^{37}$ \unitL) and have no correlation with orbital phase \citep{Stella1986}. More recently, it has been recognised that BeXRB behaviour is less dichotomous with many outbursts not easily classifiable in this current framework \citep{Kretschmar2012}. Additionally, a subset of BeXRBs, displaying persistent low level X-ray luminosity of around $10^{35-36}$ \unitL (e.g. \citealt{Bartlett2013}) with no evidence of outbursts, have been discovered. For a full review on BeXRBs and their properties see \cite{Reig2011}.

The Small Magellanic Cloud (SMC) plays host to a large number of HMXBs all of which, with the exception of SMC X-1, are all BeXRBs \citep{Coe2015}. The reason for this is not fully understood but is thought to be in part due to its recent star formation history and its low metalicity \citep{Dray2006}. Its small angular extent and its location in the sky, out of the plane of the Galaxy and so reasonably unaffected by attenuation by the Galactic interstellar medium, makes the SMC an ideal region for population studies of BeXRBs.

\subsection{\sxp}\label{intro:sxp}

The subject of this paper is \sxp, a 7.92~s pulsar in the SMC. First discovered by \citet{Corbet2008} with \rxte, it was subject to further scrutiny by \citet{Coe2009}. The lack of imaging capabilities of \rxte\ meant it was not possible to localise \sxp\ any further than approximately one quarter of the SMC. However, they determined that the likely optical counterpart to the pulsar was the early type star AzV285, based on a single \swift\ X-ray source identified in a total of nine 1~ks Target of Opportunity (ToO) observations, which covered the \rxte~FWHM zone. These observations were made approximately one month after the initial \rxte\ discovery. Whilst the lack of pulsations detected made it impossible to definitively say that this source and the \rxte\ source were one and the same, \citealt{Coe2009} made a compelling probability argument for this based on its proximity to AzV285, which showed clear evidence for binary modulation in the expected range \citep{Corbet1984}.

Using \cxo, \citet{Israel2013} reported the detection of  a 7.9 s pulsar in data from September of that year. Although the source was not consistent in position with AzV285, we regard this X-ray source as identical to SXP 7.92, based on the agreement in the pulse period. The same source was also serendipitously observed in the \integral/\xmm\ monitoring campaign of SXP\,5.05 \citep{Coe2015}. Here we discuss this recent X-ray data in detail along with previous X-ray observations and the long term optical light curve of the newly determined counterpart, identified as a variable $V=16.0$~mag object by \citet{Israel2013}.

\section{Observations and Data Reduction}\label{obs}

The region of the SMC occupied by the position reported by \citet{Israel2013} has been studied extensively in both the optical and in X-rays for the last $\sim18$~years. Here we outline the data presented in this paper in approximately chronological order, though we note that due to the serendipitous nature of the data collection, this is not strictly adhered to. All X-ray detections and non-detections for \xmm, \cxo\ and \swift\ refer to the position reported by \citet{Israel2013}

\subsection{X-ray}
\subsubsection{RXTE}\label{obs:rxte}

The \rxte\ observations used in this paper come from the regular monitoring of the SMC carried out over the period 1997-2012. The SMC was observed once or twice a week and the X-ray activity of the system determined from timing analysis. See \citet{Laycock2005} and \citet{Galache2008} for detailed reports on this technique. A summary of most of the years of observing \sxp\ was presented in \citet{Coe2009}, here we report here on observations which extend the published record on \sxp\ by three further years - but with no further \rxte\ detections in that time. As discussed in \citeauthor{Laycock2005} and \citeauthor{Galache2008}, the quality of any single observation depends upon the significance of the detected period combined with the collimator response to the source. We remove any period detections with a significance less than 99\%, and a collimator response less than 0.2.

\subsubsection{Chandra}\label{obs:Chandra}

\cxo\ has observed the position of \sxp\ four times, once in 2002 and three times in 2013, see Table \ref{tab:xray}. The observations were taken with the ACIS instrument in imaging (Timed Exposure) mode, with readout times of approximately 3.2~s. The data were reprocessed and analysed following standard procedures with the \textsc{ciao} software package version 4.8 and the calibration data base \textsc{caldb} 4.7. Source spectra and event lists were extracted from regions of $\sim$6--8~arcsec size (depending on the off-axis distance), while for the background we used annuli with internal/external radii of 15/20~arcsec. The spectra, the redistribution matrices, and the response files were created with \textsc{specextract}. For the timing analysis, we applied the Solar system barycentre correction to the photon arrival times using \textsc{axbary}. The detection algorithm used is based on the one described by \citet{Israel1996}.

\sxp\ was not detected in the 2002 observation 2948. We set a 3$\sigma$ upper limit on the source count rate of $\sim$$2.2\times10^{-3}$~counts s$^{-1}$ (0.3--8~keV). The best fit model parameters from Section \ref{sect:xray_spect} were used along with \cxo\ redistribution matrix and response files, to translate this into the 0.2--12.0~keV flux limit included in Table \ref{tab:xray}, using \textsc{xspec}'s \textsc{fakeit} task.

The discovery of 7.9-s period pulsations from \sxp\ is part of the results obtained from a larger project, the \emph{Chandra ACIS Timing Survey at Brera And Rome astronomical observatories} (CATS@BAR; \citealt{Israel2016}). This is a Fourier-transform-based systematic search for new pulsating sources in the \cxo\ Advanced CCD Imaging Spectrometer public archive and carried out in an automatic fashion.

\subsubsection{XMM-Newton}\label{obs:xmm}

\xmm\ has observed the position of \sxp\ on 12 occasions since 2000, including the three observations included in \citet{Coe2015}. Details of the observations are found in Table \ref{tab:xray}. The pre-2013 (MJD $\sim$ 56290), 3$\sigma$ upper limits were obtained using the FLIX upper limit server\footnote{\url{http://www.ledas.ac.uk/flix/flix_dr5.html}}, which makes use of the algorithm described by \citet{Carrera2007}. The appropriate canned redistribution matrices and response files were then used, along with the best fit model parameters in Section \ref{sect:xray_spect}, to correct these flux values from the assumed input spectrum ( \nh=$3\times10^{20}$~\cmsq~and $\Gamma=1.7$) to that of \sxp\ using \textsc{xspec}'s \textsc{fakeit}. The flux limits given in Table \ref{tab:xray} are the most sensitive obtained from any of the EPIC instruments onboard \xmm\ during the observation.

The three most recent \xmm\ observations of \sxp, discussed extensively in this paper, are from the 2013 targeted monitoring campaign of SXP\,5.05. For details of the general data reduction steps of these observations see \citet{Coe2015}. \sxp\ is located $\sim5.5$~arcminutes to the north east of SXP\,5.05 and fell into the field of view of all observations, though we note that, due to the observing strategy employed, \sxp\ was not visible to all three instrument in every observation: In observations 0700580101 and 0700580601 (the first and final observations), the EPIC-pn detector was in Small Window Mode to avoid pile up, the reduced field of view of this mode meant that \sxp\ was not visible to this instrument. In observation 0700580401 (the second observation) the EPIC-pn detector was in full frame mode, and so \sxp\ was visible, but the source fell on a chip gap in the two EPIC MOS detectors.

For observations where the source was detected (observations 0700580401 and 0700580601), light curves and spectra were extracted from the relevant instruments. The extraction regions were defined by the \textsc{xmmsas} task \textsc{eregionanalyse}, which calculates the optimum extraction region of a source by maximising the signal to noise ratio. This resulted in extraction radii ranging from 35 to 45 arcseconds. For the EPIC-pn light curves of \sxp\, ``Single'' and ``double'' pixel events were selected with quality flag (FLAG=0), i.e. all bad pixels and columns were disregarded.  For the EPIC-MOS light curves,  ``single'' to ``quadruple'' (PATTERN$<=$12) pixel events were selected with quality flag \#XMMEA\_EM. Photon arrival times were converted to barycentric dynamical time, centred at the solar system barycenter, using the SAS task \textsc{barycen}. Background light curves were extracted from a 90 arcsecond region on a neighbouring CCD free of sources. The background subtraction was performed using the \textsc{epiclccorr} task which also corrects for bad pixels, vignetting and quantum efficiency. Spectra were extracted using the same regions as the light curves. The area of source and background regions were calculated using the \textsc{backscal} task and response matrix files were created for each source using the task \textsc{rmfgen} and \textsc{arfgen}.

The source was not detected in observation 0700580101. Again, redistribution and response files were created using the task \textsc{rmfgen} and \textsc{arfgen} and an upper limit was derived by generating a sensitivity map using the \textsc{esensmap} task. As previously, \textsc{xspec}'s \textsc{fakeit} task was used to simulate the best fit spectrum to \sxp\ with this count rate to obtain an estimate on the upper limit of the source flux at this time.

\begin{figure*}
\centering
\includegraphics[width=0.85\textwidth]{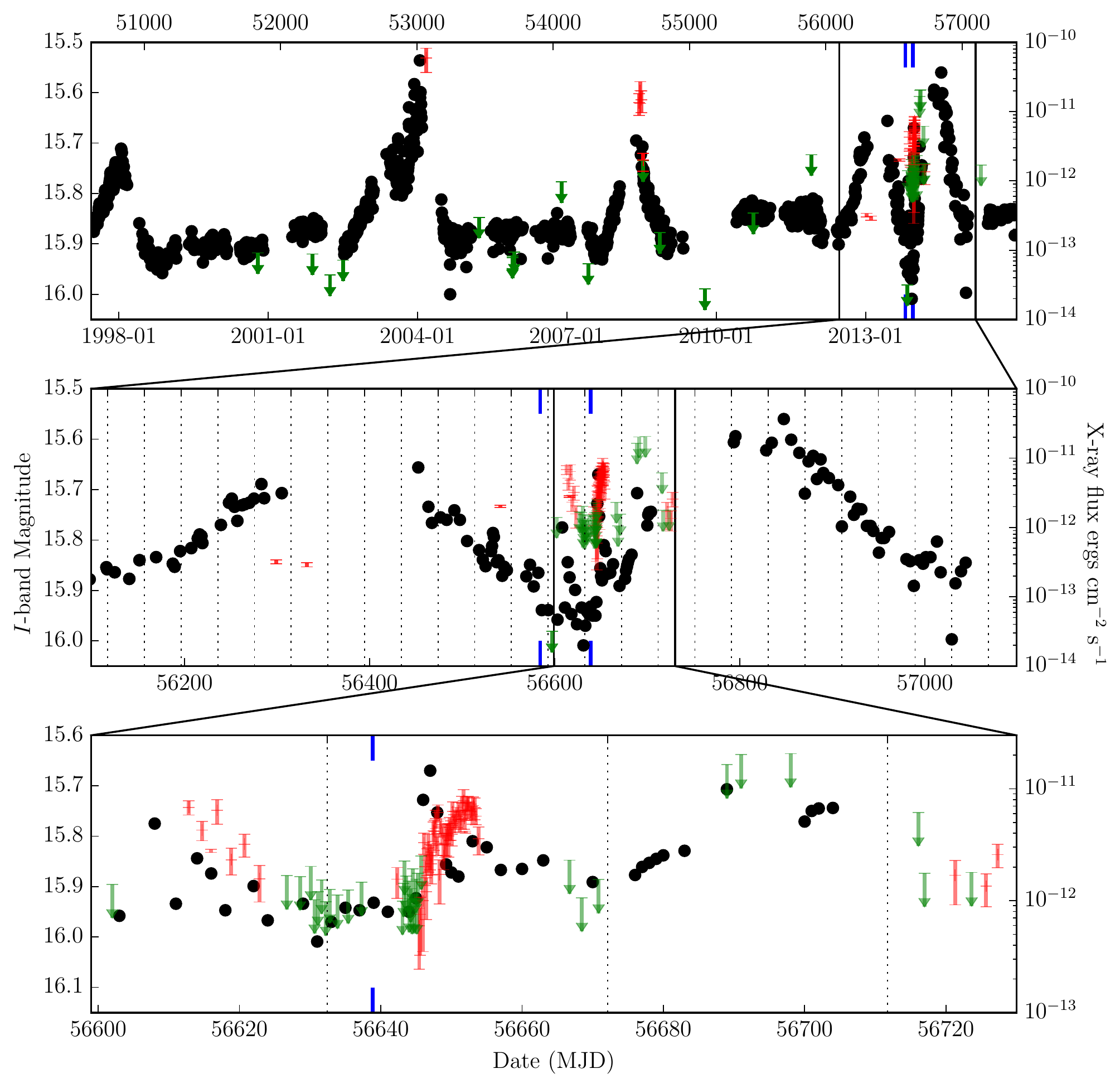}
\caption{Light curve of the optical counterpart of \sxp. The green arrows mark the $3\sigma$ X-ray upper limits from \cxo\, \xmm\ and \swift, the red points mark the X-ray detections. The epochs of the optical spectra are marked by the blue lines. The top panel shows the entire $\sim$18 year light curve, and the lower panels subsequent small subsets of the most recent optical outburst. The dotted lines represent the orbital period proposed in Section \ref{sect:Porb}. An X-ray flux of $2.3\times10^{-12}$ \unitF\ corresponds to a luminosity of $10^{36}$ \unitL\ at the distance of the SMC (62.1~kpc; \citealt{Graczyk2014})}\label{fig:ogle}
\end{figure*}
 
\subsubsection{Swift}\label{obs:swift}

\sxp\ has been observed with XRT onboard \swift\ on 129 occasions since 2006 November, of which 118 observations were in photon counting (pc) mode and 11 were in window timing (wt) mode. The majority of these observations were targeted at SXP\,5.05 and so are concentrated around the 2013 outburst. We used the \swift\ XRT data products generator\footnote{\url{http://www.swift.ac.uk/user_objects/}} \citep{Evans2007,Evans2009} to obtain background subtracted count rates or upper limits of \sxp\ in these observations.  Of the 11 XRT-wt observations, 9 were declared ``unsafe'' as no centroid could be determined. These observations were not included in our analysis. The XRT-pc count rates for each observation were converted to source flux values using the \swift\ canned response matrix and best fit model to the source spectrum with \textsc{xspec}'s \textsc{fakeit}. The wt observations were not treated in this manner, as the 1 dimensional observations are dominated by SXP\,5.05.

The XRT-wt and pc observations in which \sxp\ was detected were downloaded and processed with \textsf{xrtpipeline} with the position of \sxp\ specified. Event files were extracted with a 20 arcsec circular region and a barycentric correction was applied using the mission independent tool \textsf{barycorr}. Source lightcurves were created and searched for periods, however none were detected due to a combination of low count rates ($\sim0.06$~counts s$^{-1}$ at peak for the XRT-pc observations), short exposures times ($<5$~ks with many $<1$~ks) and, in the case of the XRT-pc data, low time resolution (2.506~s).

A combined spectrum of all the positive detections, along with the associated redistribution matrix, response and background files was created with the XRT data products generator and is included in Section \ref{sect:xray_spect}.

\subsection{Optical}
\subsubsection{Photometry}\label{obs:optphot}

Data from the Optical Gravitational Lensing Experiment (OGLE)  Phase III and IV  \citep{Udalski2003,Udalski2015} are used to investigate the long-term behaviour of the identified optical counterpart of \sxp. OGLE has been regularly monitoring this object since 2000 with the 1.3~m Warsaw telescope at Las Campanas Observatory, Chile, equipped with two generations of CCD camera: an eight chip 64 Mpixel mosaic (OGLE-III) and a 32-chip 256 Mpixel mosaic (OGLE-IV). Observations were collected in the standard $I$-band. 

\subsubsection{Spectroscopy}\label{obs:optspec}

Optical spectra of the optical counterpart of \sxp\ were taken on the nights of 2013 October 18th and 2013 December 1st. The first observations were taken with the ESO Faint Object Spectrograph (EFOSC2, \citealt{Buzzoni1984}) mounted at the Nasmyth B focus of the 3.6m New Technology Telescope (NTT), La Silla, Chile.  The instrument was in longslit mode with a slit width of 1.0 arcsec and instrument binning $2\times2$. Grism \#20 was used to obtain a spectrum around the \Halp\ region (6563~\AA{}). Grism \#20 is a Volume-Phase Holographic grism with 1070 lines~pixel$^{-1}$, blazed at 6597\AA{}. This setup leads to a dispersion of $\sim1.0$ \AA{}~pixel$^{-1}$ and a resolution $\sim7$~\AA{} (as determined from the FWHM of the emission lines in the comparison spectrum). The spectra were extracted using standard procedures (bias and background subtraction, flat fielding and wavelength calibration) using the standard \textsf{IRAF} packages. A He-Ar lamp spectrum was acquired during the same night as the target and used for wavelength calibration.

The later observation was taken with Robert Stobie Spectrograph (RSS, \citealt{Burgh2003}) on the Southern African Large Telescope (SALT, \citealt{Buckley2006}) at the South African Astronomical Observatory. Grating PG2300 was used with a grating angle of 30.5 degrees, a slit width of 1.5 arcsec and $2\times4$ (spectral $\times$ spatial) binning. The setup described results in a dispersion of $\sim0.25$ \AA{}~pixel$^{-1}$ at the central wavelength, 4600\AA{}, with chip gaps of $\sim 16$~\AA{} centred at 4221~\AA{} and 4587~\AA{}. 

The data were reduced with the python based reduction pipeline and software package for SALT, \textsc{PySALT} \citep{Crawford2010}.  Wavelength calibration was achieved using comparison spectra of Copper and Argon lamps, taken immediately after the observation with the same instrument configuration. One dimensional, background subtracted spectra were then extracted and normalised using the standard \textsc{IRAF} packages.

\section{Results and Analysis}\label{sect:results}
\subsection{Overview}

Figure \ref{fig:ogle} shows the $\sim$18 years of OGLE $I$-band photometry, as well as an overview of the long term X-ray activity of the source, and summarises the optical and X-ray data presented in this paper. \sxp\ was detected on two occasions by \rxte: MJD = 53068 and MJD = 54630 -- 54645. These two detections are indicated on Fig.~\ref{fig:ogle} and clearly coincide with the peaks of optical outbursts. Though these optical outbursts each lasted 1 -- 2 years, and the \rxte\ coverage was extensive throughout these periods, \sxp\ was only detected on those two occasions. We cannot definitively draw any conclusion as to the \emph{absolute} flux of the source during these intervals, but any source with \lx$\sim5$\ergs{35} and a pulsed fraction \textgreater0.3 would have been detected in these observations. 

\subsection{X-ray Position and Timing}\label{results:timing}

\citet{Israel2013} reported the detection of an X-ray transient in \cxo\ observation 15504 (MJD = 56541) at a position RA(J2000)=00:57:58.4, Dec(J2000)=-72:22:29.5 with a $1\sigma$ uncertainty of 1.5\arcsec. The source was subsequently searched for periodic signals by means of an ad hoc detection algorithm and showed a large and highly significant (confidence level above 12$\sigma$) signal at a period of about 7.9~s (Nyquist period at 6.3~s) in its power spectrum, confirming it to be \sxp\ and the same source as seen by \rxte. 

In order to reject the hypothesis of a spurious result, the signal was automatically checked by means of the \textsf{dither\_region ciao} task\footnote{\url{http://cxc.harvard.edu/ciao/ahelp/dither_region.html}} and confirmed to be intrinsic to the source. A more reliable and accurate period was derived by performing a phase-fitting analysis by dividing the light curve in five sub-intervals. We obtained a best period value of $P=$7.91807$\pm$.00001~s and a pulsed fraction (defined as the semi-amplitude of the sinusoid divided by the average count rate) of 50$\pm$2\%. The pulse shape is almost sinusoidal, within the relatively poor time resolution of the data (intrinsic ACIS binning time of 3.14~s).   

A search of the \cxo\ archive revealed that \sxp\ is detected in two previous \cxo\ observations, albeit at a fainter level (see Table \ref{tab:xray}). Shortly after, \sxp\ was detected in two of the three \xmm\ observations of \citet{Coe2015} (see Table \ref{tab:xray}) at positions RA(J2000)=00:57:58.7, Dec(J2000)=-72:22:28.0 and RA(J2000)=00:57:58.7, Dec(2000)=-72:22:28.4 respectively. This position is $\sim$20.6 arcminutes to the south-east of the originally reported position and conclusively rules out AzV285 as the optical counterpart to \sxp.

The \xmm\ light curves were searched for periods using the numerical implementation of the Lomb--Scargle method \citep{Lomb1976,Scargle1982,Press1989}. 
Figure \begin{figure}
\centering
\includegraphics[width=0.48\textwidth]{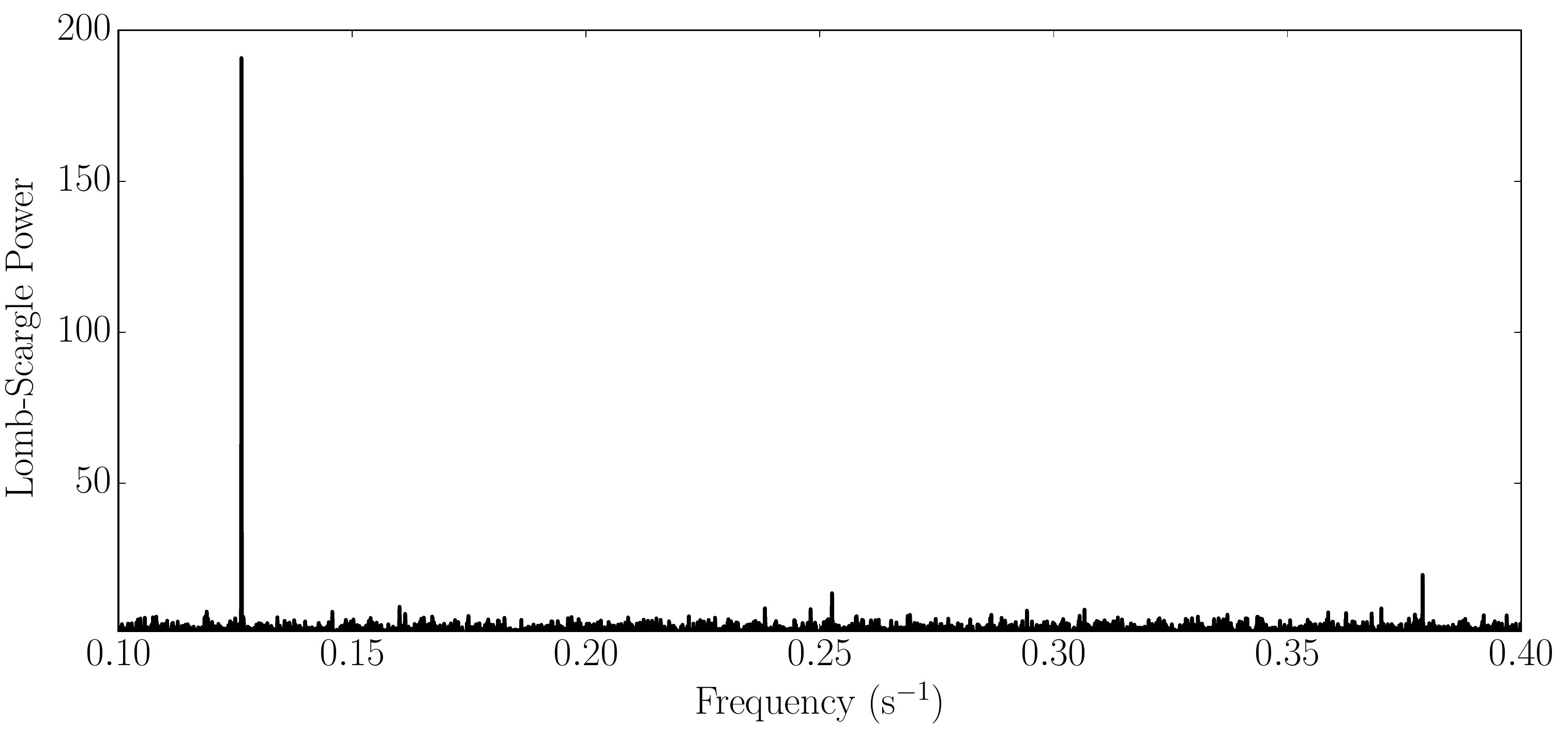}
\caption{Example periodogram of \sxp. Figure shows the Lomb-Scargle periodogram of the 0.2--10.0~keV EPIC-pn light curve of \sxp\ from observation 0700580401.}\label{fig:xmm_lombscar}
\end{figure} \ref{fig:xmm_lombscar} shows a resulting periodogram from this search, generated from the EPIC-pn light curve of observation 0700580401. A strong signal is detected in both observation 0700580401 and observation 0700580601 at a periods 7.917~s.

The errors on the detected periods were determined via a \emph{bootstrap-with-replacement} method: 1000 artificial light curves were generated for each observation by sampling with replacement from the original light curve. Lomb-Scargle analysis was performed on each of these light curves in an identical manner to that of the original light curve and the period determined. The distributions of the resulting periods detected are well characterised by Gaussian functions with mean 7.9174 and 7.9167 and standard deviation $1\times10^{-4}\text{and}2\times10^{-4}$ respectively. Thus we determine the period of \sxp\ to be $7.9174\pm.0001$~s as of MJD = 56616 and $7.9167\pm.0002$~s as of  MJD = 56652.  \begin{figure}
\centering
\includegraphics[width=0.32\textwidth]{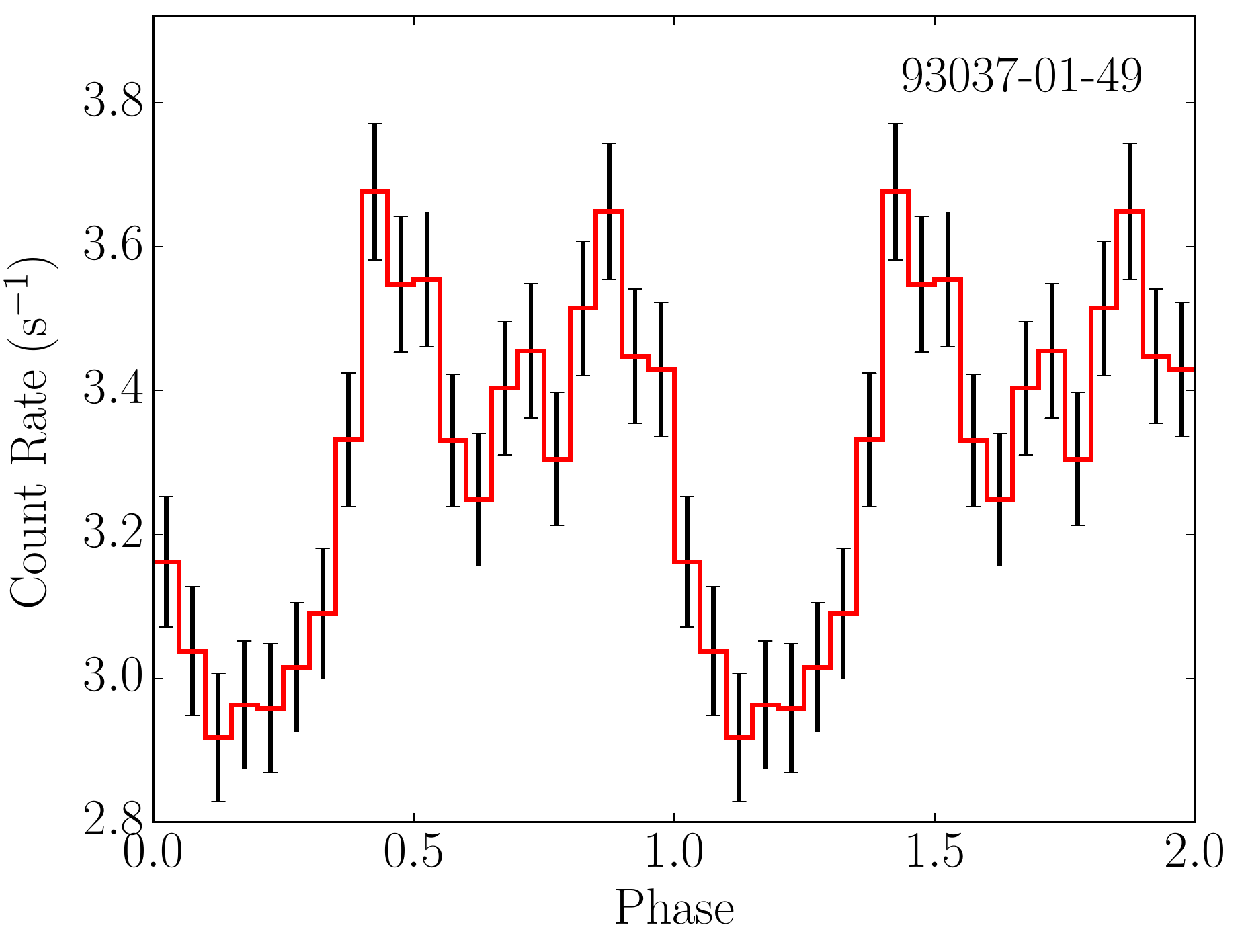}
\includegraphics[width=0.32\textwidth]{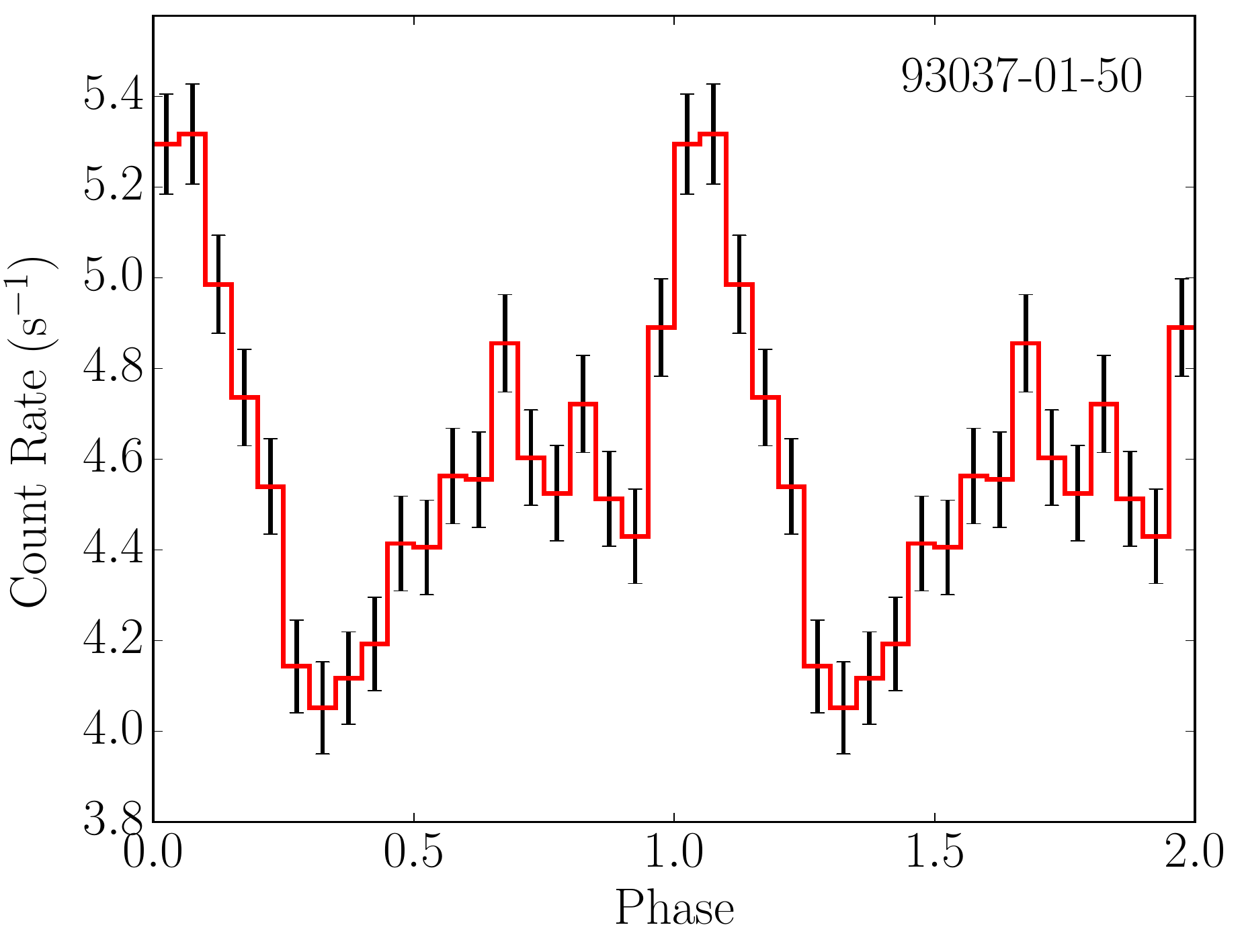}
\includegraphics[width=0.32\textwidth]{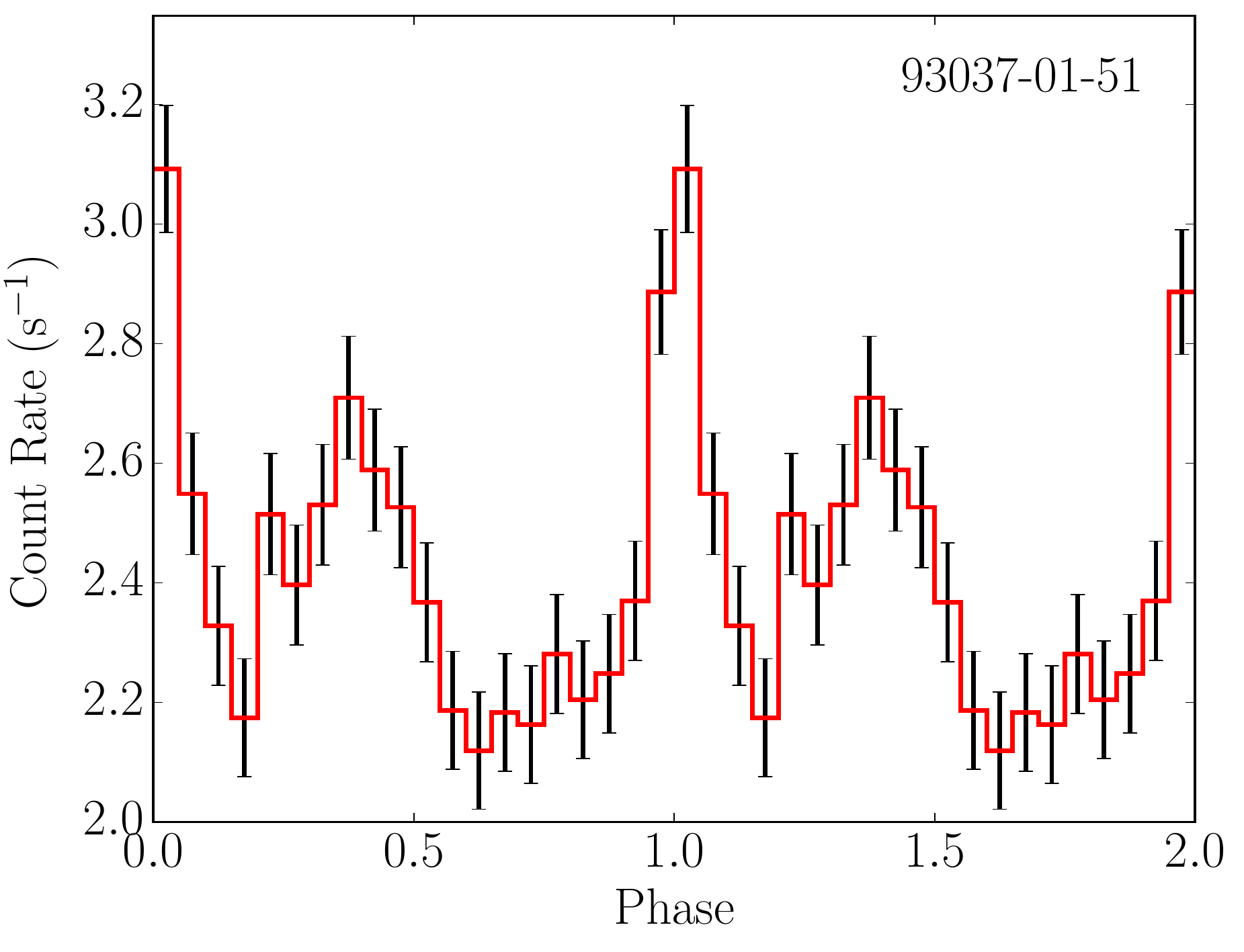}
\includegraphics[width=0.32\textwidth]{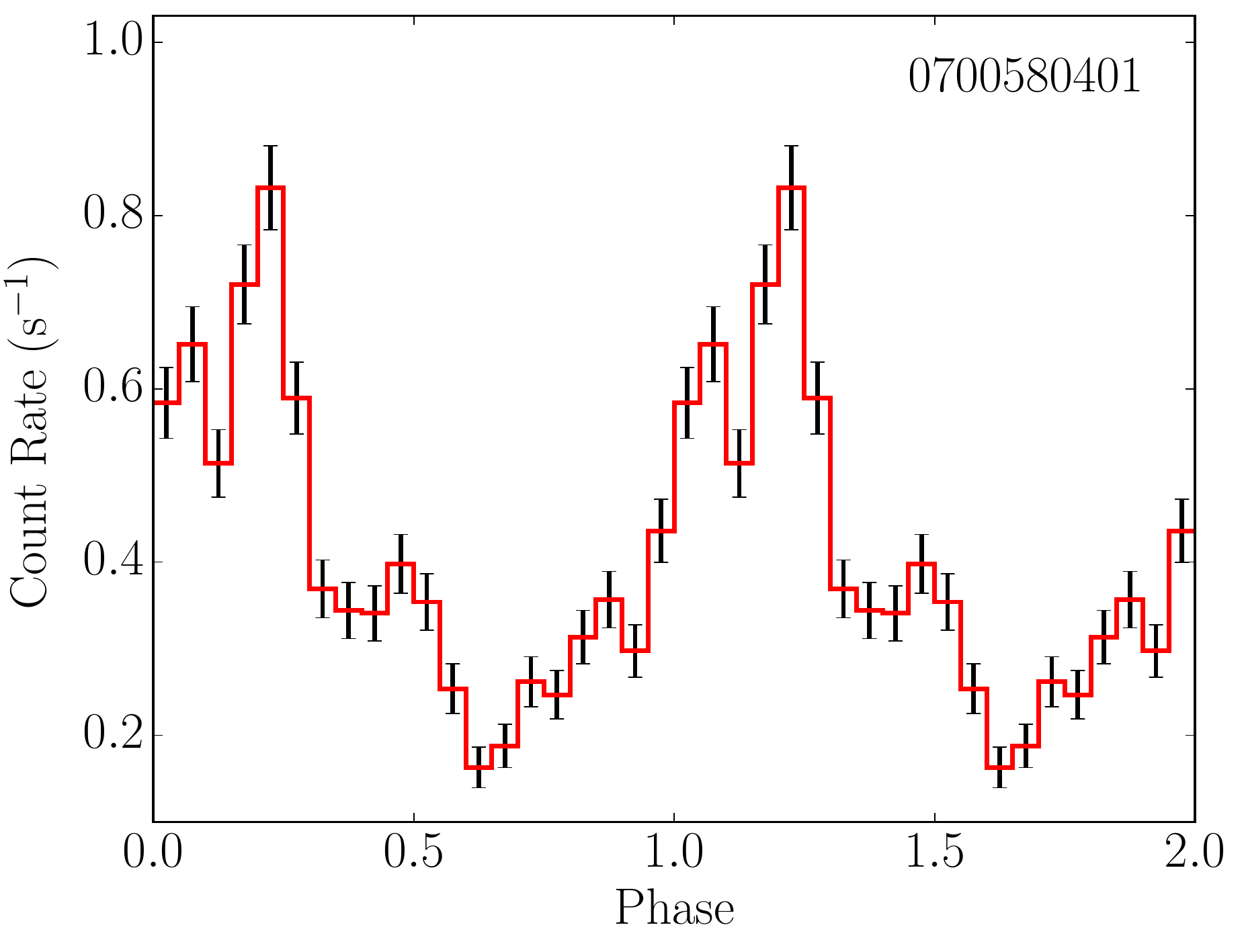}
\caption{\rxte\ and \xmm\ pulse profiles of \sxp, so chosen as these instruments have the best timing resolution and so offer the best insight into the variation in the shape of the pulse profile. The count rate for the \rxte\ profiles (top three panels) represent the count rate for the entire observation, i.e. across the whole field of view and without any background filtering applied. As such, these count rate values are inclusive of the X-ray background and may contain contributions from other sources in the field of view.}\label{fig:pprofiles}
\end{figure}
Table\begin{table}
\centering
\caption{Summary of the X-ray timing results presented in this paper. As discussed in the text, the lack of imaging capabilities of \rxte\ means that it is impossible to determine the background level and hence the absolute flux and pulsed fraction of the source in these observations.}\label{tab:xray_timing} 
\begin{tabular}{lccc}
\hline
Observation 		& Date			& Pulsed Period		& Pulsed Fraction		 \\
\hline
90086-01-01		& 2004-03-04		& 7.9$\pm$0.1			& - 	\\
93037-01-49		& 2008-06-13		& 7.918$\pm$0.002		& -  	\\
93037-01-50		& 2008-06-20		& 7.9206$\pm$0.0001	& -	\\
93037-01-51		& 2008-06-28		& 7.92$\pm$0.05		& -	\\	
15504			& 2013-09-06		& 7.91807$\pm$.00001	& 0.50$\pm$0.02		\\
0700580401		& 2013-11-20		& 7.9173$\pm$0.0001	& 0.60$\pm$0.02		\\
0700580601		& 2013-12-26	 	&7.9167$\pm$0.0002	& 0.34$\pm$0.01		\\
\hline
\end{tabular}\end{table} \ref{tab:xray_timing} summarises these results. 

Figure \ref{fig:pprofiles} shows the \rxte\ and \xmm\ pulse profiles (folded light curves) for a subset of observations, so chosen because (a) they demonstrate the range of pulse profiles shapes shown by this sources and (b) because these instruments and modes have high timing resolution (i.e. significantly less than the pulse period, unlike \cxo\ or the EPIC-MOS instruments onboard \xmm.) Each light curve has been folded on the best period detected between 7.0~s and 8.5~s in each observation, with the phase = 0 point defined by the start time in observation 93037-01-49 (though we note that we have made no correction to take into account the changing pulse period between the Phase = 0 point and each individual observation). The pulse profiles all appear to have a common double, possibly triple, peaked structure, varying from symmetric to asymmetric. The triple peak shape appears to be more prominent in the final, most recent pulse profile, but making such inferences across instruments with different resolutions and different signal to noise data should be done with caution. The shape and evolution of the \rxte\ pulse profiles are discussed at length in \citet{Coe2009}, who conclude they are likely caused by changes in accretion mode and geometry due to variations in the mass accretion rate. The pulsed fractions of the \cxo\ and \xmm\ light curves, included in Table \ref{tab:xray_timing}, were calculated by integrating over the pulsed profile and computing the fraction of the profile that pulsed vs. the total flux. 

\subsection{X-ray Spectra}\label{sect:xray_spect}

\begin{figure}
\centering
\hspace{-30pt}
\includegraphics[width=0.5\textwidth]{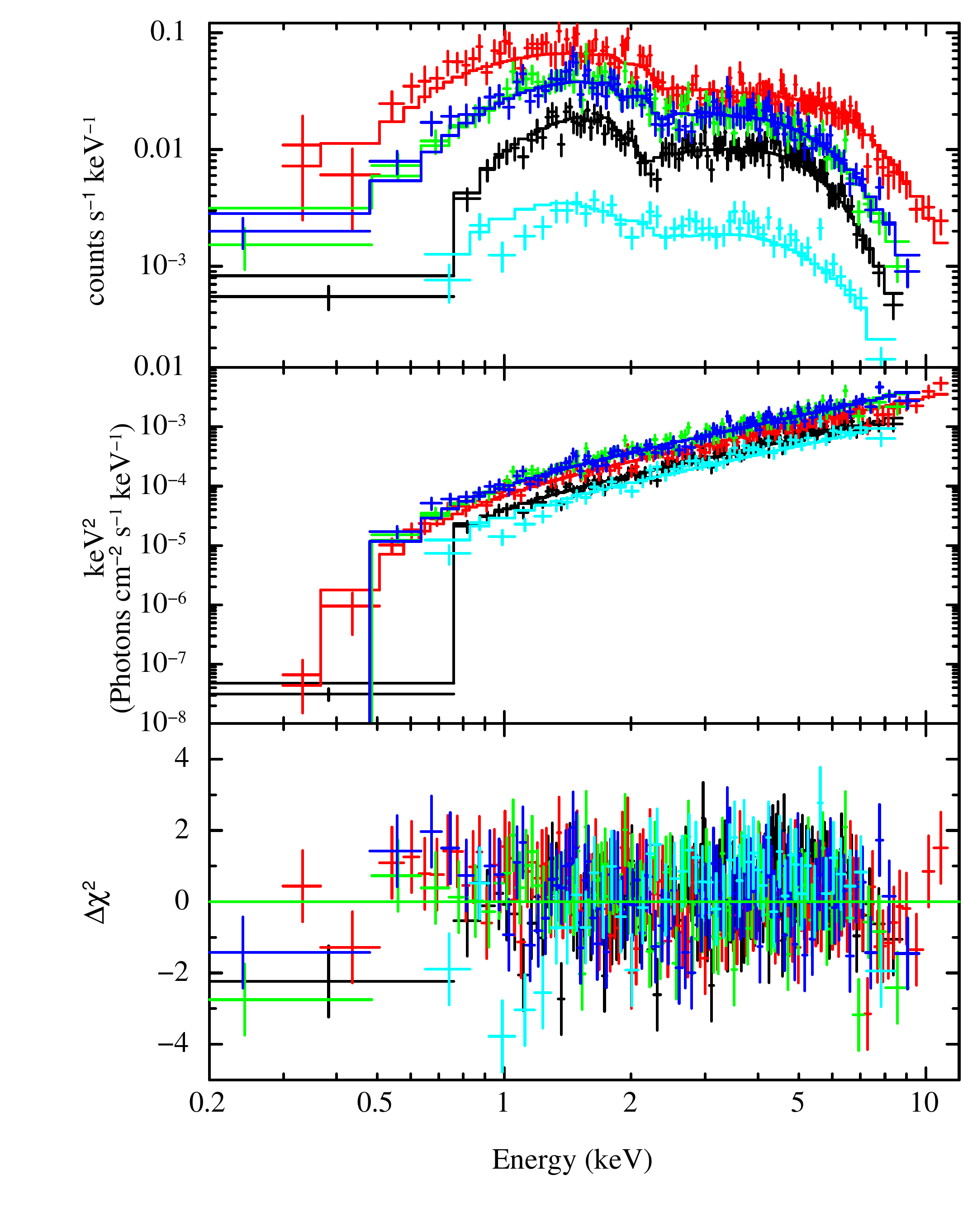}
\caption{The 0.2--12.0~keV spectrum of \sxp\ from the various X-ray telescopes. From top to bottom in the top panel the \xmm\ EPIC-pn (red), \xmm\ EPIC-MOS1 (green) and EPIC-MOS2 (blue), the \cxo\ ACIS-I (black) and the \swift\ combined XRT (cyan) spectrum. The top two panels show the background subtracted spectrum with the best fit \emph{phabs*vphabs*powerlaw} model, fit to all observations simultaneously. The top panel shows this convolved with the relevant instrumental responses, the middle panel shows the same in $\nu-F\nu$ form. The bottom panel shows the residuals.}\label{fig:spec}
\end{figure}

The spectral analysis reported here was performed using \textsf{XSPEC} \citep{Arnaud1996} version 12.9.0i. The spectra of the recent \cxo\ and \xmm\ observations (i.e. the recent period of activity) were fit both individually, to see if there was any evidence for variability, and simultaneously with the \swift\ combined spectrum, to better constrain the model parameters. The spectra were fit with an absorbed power law (\emph{phabs*vphabs*powerlaw} in \textsf{XSPEC}); the typical model for BeXRBs \citep{Haberl2008}. The photoelectric absorption was split into two components, a Galactic component, $N_{\rm H,Gal}$ ( the \emph{phabs} component), to account for Galactic foreground extinction, fixed to $6\times10^{20}$ \cmsq\ \citep{Dickey1990} with Solar abundances from \citet{Wilms2000}, and an intrinsic absorption, $N_{\rm H,i}$ to account for attenuation local to the source and the SMC with elemental abundances of 0.2 solar (for elements heavier than He; the \emph{vphabs} component). Table \begin{table}
\centering
\caption{Best fit parameters for the \emph{phabs*vphabs*powerlaw} model fit to the data, both to individual observations and simultaneously. The \emph{phabs} component corresponds to $N_{\rm H,Gal}$, and was fixed at $6\times10^{20}$ \cmsq. Errors, where reported, are the 90\% confidence intervals. ``dof'' refers to the degrees of freedom of the fit.}\label{tab:xray_spec}
{\scriptsize \begin{tabular}{llccr}
\hline\noalign{\smallskip}	
Observation 				& Instrument 		&  $N_{\rm H,i}$  					& $\Gamma$				 			& $\chi^2$/dof 			\\\noalign{\smallskip}
						&				& $\times10^{22}$~\cmsq				&									&					 \\\noalign{\smallskip}
\hline\noalign{\smallskip}
15504 					& ACIS-I 			& $0.4_{-0.2}^{+0.3}$ 				& $0.40_{-0.08}^{+0.09}$ 					& 100.0/118			  \\\noalign{\smallskip}
0700580401				& EPIC-pn 		& $0.29_{-0.08}^{+0.09}$ 				& $0.57\pm0.06$ 						& 100.9/122			   \\\noalign{\smallskip}
\multirow{2}{*}{0700580601} 	& EPIC-MOS1 		& \multirow{2}{*}{$0.31_{-0.08}^{+0.09}$} 	& \multirow{2}{*}{$0.55\pm0.05$}			& \multirow{2}{*}{186.5/185} \\\noalign{\smallskip}
						& EPIC-MOS2 		& 								& 									& 					    \\\noalign{\smallskip}
\hline\noalign{\smallskip}	
\multicolumn{2}{c}{Simultaneous}				& $0.32_{-0.06}^{+0.07}$			         & $0.49\pm0.03$						& 506.4/485			\\\noalign{\smallskip}
\hline
\end{tabular}}\newline
\end{table}\ref{tab:xray_spec} shows the model parameters for both the individual and simultaneous fits to the data. All the observations are well fit by this simple model with no evidence for variability in the spectral shape between the observations (i.e. the absorbing column and the photon index are consistent with being constant) despite the variability in source flux, represented by a variable normalisation parameter in the spectral fit (see middle panel of Fig. \ref{fig:spec}).

Figure \ref{fig:spec} 
shows the five spectra, the \cxo\ ACIS-I spectrum from observation 15504 (black), The \xmm\ EPIC-pn spectrum from observation 0700580401 (red), the EPIC-MOS1 and MOS2 spectra from observation 0700580601 (green and blue respectively) and the combined \swift\ XRT spectrum (cyan), along with the simultaneous model fit. \begin{figure*}
\centering
\includegraphics[width=1.0\textwidth]{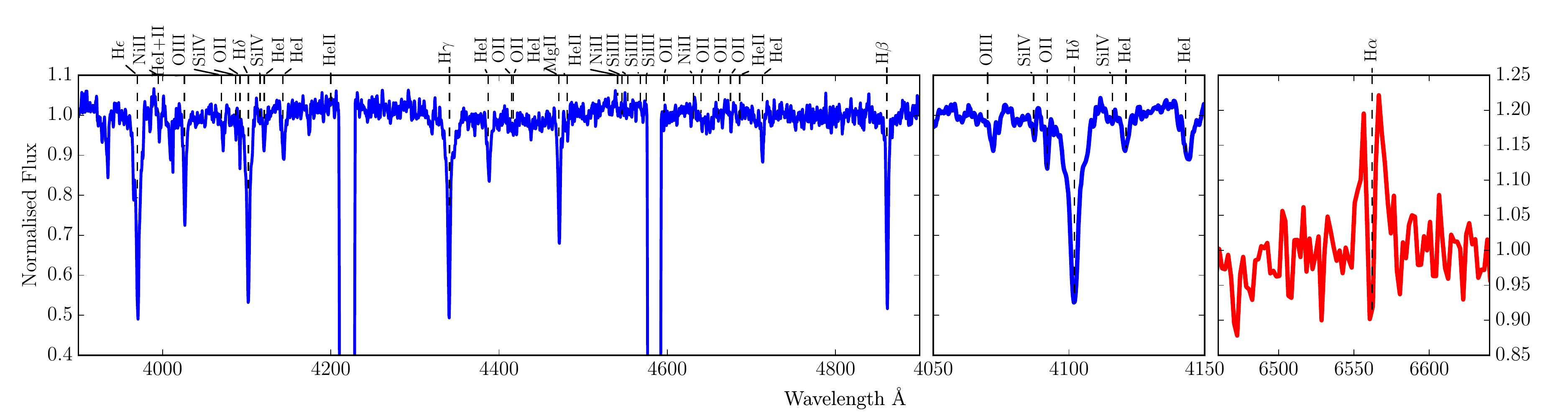}
\caption{Left hand panel shows the SALT spectrum of \sxp\ in the wavelength range 3900--4900\AA{}. The spectrum has been normalised to removed the continuum and redshift corrected by -158~km~s$^{-1}$. Atomic transitions relevant to spectral classification have been marked. The middle panel shows a subset of the data that is discussed explicitly in the text to facilitate identification of the weaker metal lines. The right hand panel shows the EFOSC2 \Halp\ spectrum of \sxp\ . Again, the spectrum has been normalised to removed the continuum and redshift corrected by -158~km~s$^{-1}$. The broken line at 6563\AA{} is the rest wavelength of \Halp\ and is centred on the middle depression}\label{fig:opt_spec}
\end{figure*}

\subsection{Spectral Classification}

Figure \ref{fig:opt_spec} shows the normalised SALT spectrum of the optical counterpart of \sxp\, taken on MJD = 56639 (2013-12-12). This spectrum corresponds to the second blue line in Fig. \ref{fig:ogle} and was taken just after the sharp minimum, between the two positive \xmm\ detections (MJD = 56616 and 56663). A box filter smoothing of 3 pixels has been applied\footnote{\url{http://docs.astropy.org/en/stable/convolution/index.html}} as well a redshift correction of -158\kms \citep{Richter1987} corresponding to the recessional velocity of the SMC. The absorption lines used to classify early type stars are labelled. \citet{Evans2004} present a detailed discussion of the difficulties in classifying low luminosity (i.e. sub giant) B-type stars in the SMC: It is near impossible to determine whether the absence of a metal line in a stellar spectrum is due to temperature or the result of the low luminosity and/or low metallicity environment. As such, we use the method described by \citet{Evans2004} and further implemented by \citet{Evans2006} and adopted a conservative approach in our classification.

The lack of evidence for \HeII $\lambda\lambda 4200, 4541\text{or}4686$ above the noise of the data, implies a spectral class later than B0.5. There appears to be some evidence for \SiIV\ lines, particularly at $\lambda4088$, suggesting a spectral type of B1-B2. The strength of the \SiIV $\lambda4088$ line relative to the neighbouring \OII $\lambda4097$ line in particular points to a B1.5 spectrum. The \HeII $\lambda\lambda4144/4121$ ratio is indicative of a main sequence dwarf star (i.e. luminosity class V) if we assume a spectral class B1, or a subdwarf/dwarf star (IV-V) if B2. The far right panel of Figure \ref{fig:opt_spec} shows the NTT spectrum of \sxp\ around the \Halp\ $\lambda6563$ region of the spectrum. There is unambiguous evidence of emission, justifying an `e' classification for the optical counterpart.

\citet{Zaritsky2002} quote a visual magnitude for the optical counterpart to \sxp\ of $V=15.72\pm0.03$. Assuming a distance to the SMC of $62\pm2$~kpc \citep{Graczyk2014} and using the relation of \citet{Guver2009} and an interstellar extinction value of $6\times10^{20}$~\cmsq \citep{Dickey1990} to calculate $A_V$ gives us an absolute magnitude $M_V=-3.51\pm0.07$. Comparing this value to the tables of absolute magnitudes for OeBe IV/V stars in \citeauthor{Wegner2006} (\citeyear{Wegner2006}; B0.5Ve=$-3.63$, B1Ve=$-3.35$, B1.5Ve=$-3.05$, B2Ve=$-2.72$) supports a spectral classification of B1Ve. We note, however, that our calculated absolute magnitude may be misleading as it doesn't take into consideration any local extinction (which may vary throughout the orbit) and the error doesn't take into account the depth of the SMC, estimated to be as much as $6.2\pm0.3$~kpc \citep{Haschke2012} or any local extinction. 

\subsection{Optical Period}\label{sect:Porb}

\begin{figure*}
\centering
\includegraphics[width=1.0\textwidth]{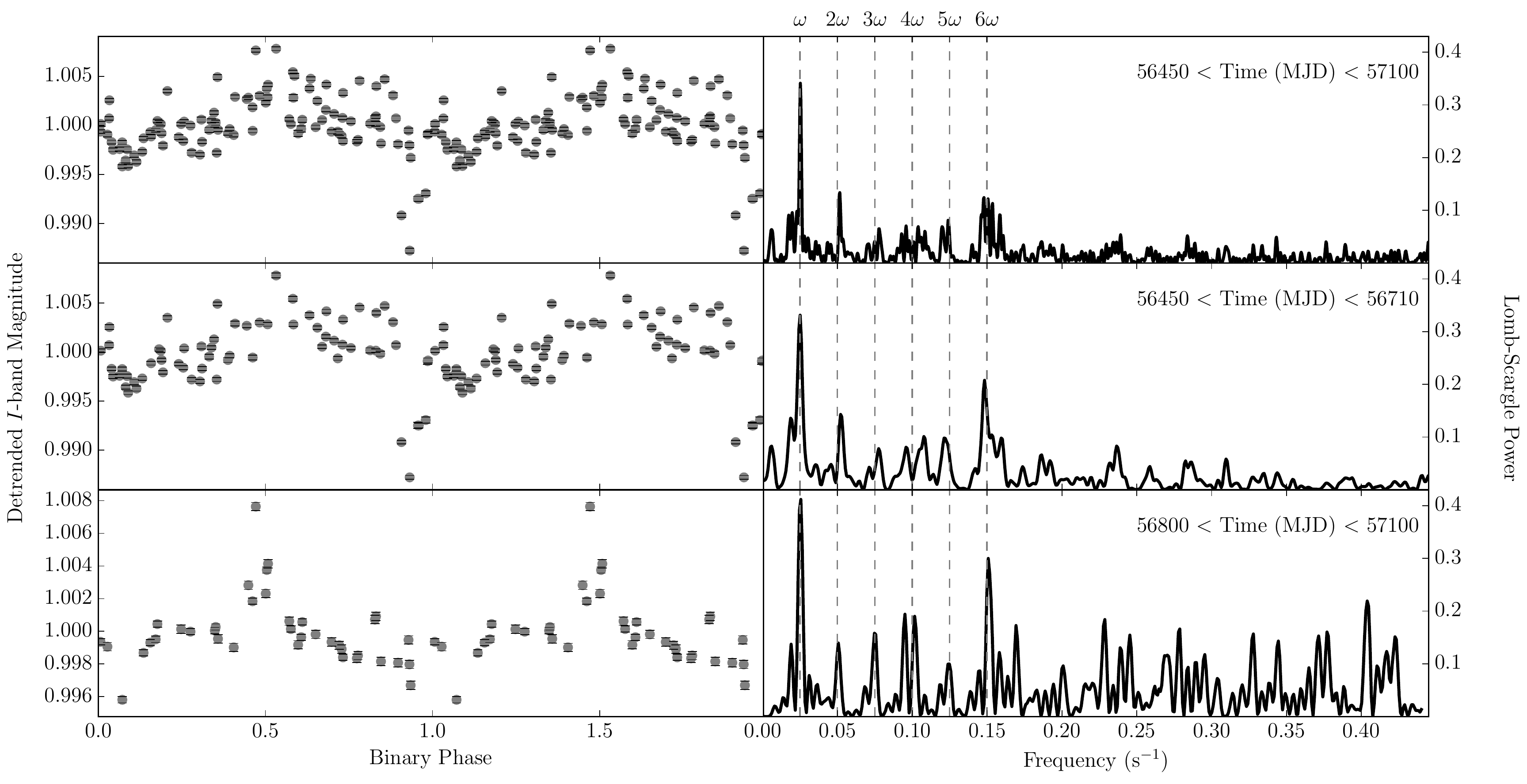}
\caption{The folded OGLE IV data and Lomb-Scargle periodograms from the periods MJD = 56500 -- 57100 (top panels), MJD 56450 -- 56700 (middle panels) and MJD 56800 -- 57100 (bottom panels) respectively. The light curves folded at the proposed binary period of 40.0~d. All light curves have the same zero phase point $T_0=56850$, and all have been been de-trended before folding. An apparent shift in the phase of the flares with respect to the dips is evident.}\label{fig:fold30}
\end{figure*}

It is clear from Figure \ref{fig:ogle} that the optical counterpart of \sxp\ spends a large fraction of its time in a quiescent state, with $V\sim15.9$, but with several major outbursts each lasting 1--2 years. Also shown in more detail in this Figure is the most recent outburst. The data from epoch MJD = 56450 -- 56710 and epoch MJD= 56800 -- 57100 were detrended with simple second and third order polynomials respectively. These intervals were searched for periodic signals, both as separate intervals and as an entire  MJD = 56450 -- 57100 period. This was performed in the same manner as the X-ray light curve: Using the numerical Lomb-Scargle method. The resulting periodograms are shown in Figure \ref{fig:fold30}. All three epochs show a strong peak at $\sim$40~d, as reported by \citet{Schmidtke2013}.

Error determination was, again, performed via bootstrapping. As with the X-ray light curve, 1000 artificial light curves were drawn from the detrended MJD = 56500 -- 57100 data set, by sampling with replacement, and were searched for periodicities in the same way as the original light curve. The resulting distribution of periods recovered is well described by a Gaussian function with mean 40.04 and standard deviation 0.32, thus we we determine the optical period of \sxp\ to be 40.0$\pm$0.3~days.

The detrended data for each of these periods were folded on the proposed binary period and are shown in Fig.~\ref{fig:fold30}. We note that there appears to be an evolution from ``flaring'' MJD = 56450 -- 56710, to dips in epoch MJD= 56800 -- 57100, offset from flare phase. The phases of these dips are indicated in Fig.~\ref{fig:ogle} by the vertical dashed lines. We note that this is not the first report of ``dips'' in the optical light curve of a BeXRB: similar sized dips were seen in the OGLE light curve of LXP\,169 \citep{Maggi2013} and SMC\,X-2 has shown a similar evolution from ``flares'' to dips \citep{Rajoelimanana2011}. This will be further discussed in Sections \ref{sect:circumstellar} and \ref{sect:discussion}.

\section{Discussion}
\subsection{The Circumstellar Environment}\label{sect:circumstellar}

\begin{figure}
\centering
\includegraphics[width=0.5\textwidth]{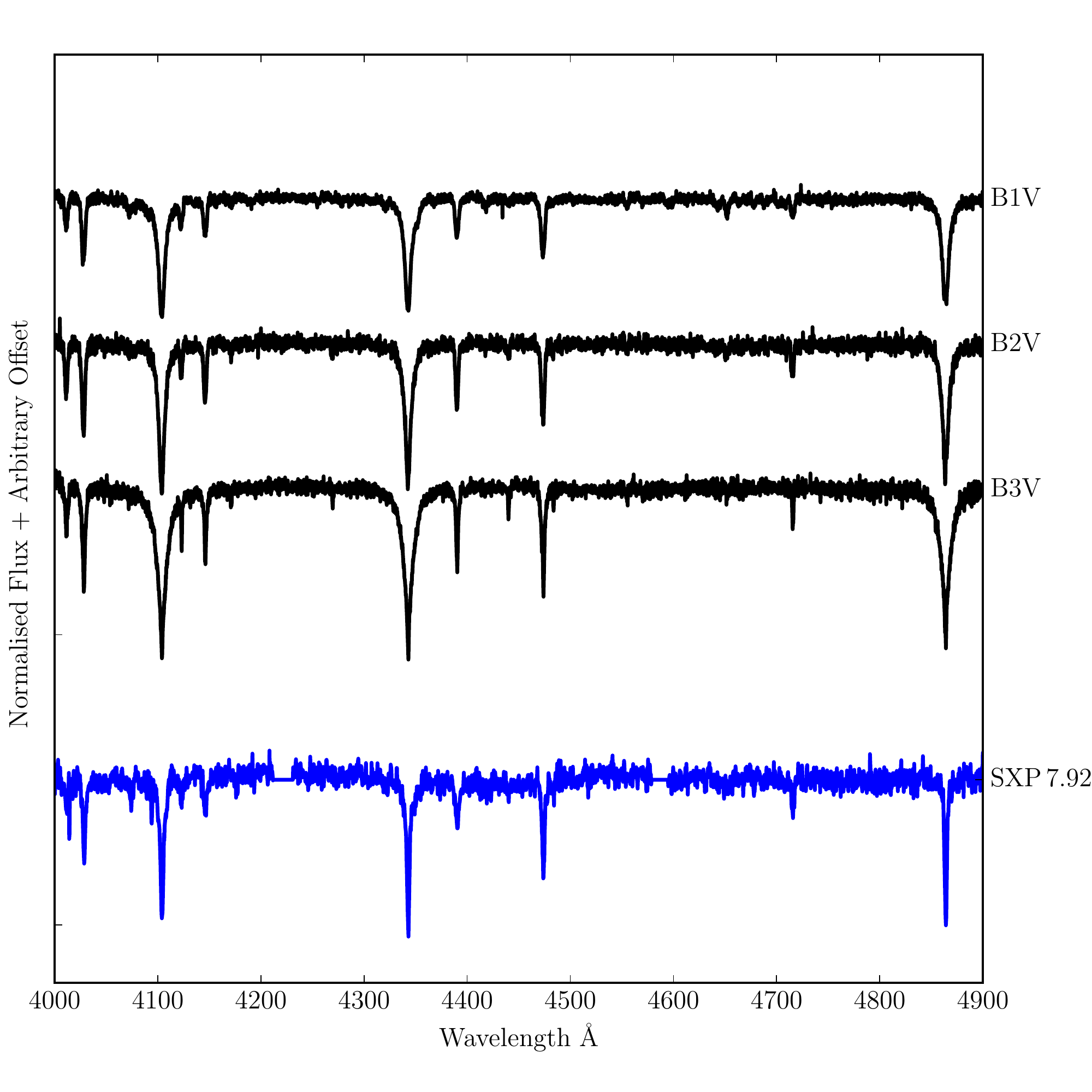}
\caption{spectrum of \sxp\ alongside those of isolated SMC B stars. Despite usually being characterised by Doppler broadened Balmer lines, the Balmer series of \sxp\ appear comparable to those in isolated B stars.}\label{fig:Bstar}
\end{figure}

Figure \ref{fig:Bstar} shows the SALT spectrum of the optical counterpart of \sxp\ alongside a selection of isolated SMC B stars. Be stars, both isolated and in BeXRBs, are characterised by Doppler broadened Balmer series due to their rapid rotation, yet the Balmer series of \sxp\ appear comparable to those in isolated B stars, suggesting \sxp\ has a low $v\sin i$. One explanation is that we are viewing the star near pole on ($\sin i\sim0$). Such a scenario, however, would lead to single peaked emission lines from the circumstellar disc whereas the \Halp\ profile is unambiguously double peaked. Double peaked \Halp\ profiles are common in BeXRBs; their shape and variability (``$V/R$ variability''; \citealt{Mclaughlin1961}) is an area of active research. These are explained by an inclined circumstellar disc with blue and red shifted emission at alternate ends. In such cases the central depression cases tends to be shallow. On occasion, the Balmer series of BeXRBs show evidence of ``infilling'', i.e. a superposition of a broad absorption line and a narrow emission feature, \sxp\ seems to show the opposite of this: emission in the wings of the line and a deep narrow core - a shell profile.

Shell lines are observed in subset ($\sim$10-20\%) of isolated Be stars and are characterised by a deep absorption core embedded in an emission line \citep{Hanuschik1996,Porter1996}. This is explained by absorption of light from the disc and the central star by the circumstellar disc itself and is a natural consequence of viewing the Be star equator or near equator on. \citet{Hanuschik1996} use the fraction of Be shell stars to calculate a half opening angle of Be star discs of 13$^\circ$ and show that these discs must be thin at the inner edge, increasing in scale height at greater radii. Along with the small half opening angle, this implies a concave (i.e. hydrostatically balanced) disc. They show that, for a narrow range of $i$, the shell phenomenon can be transient, as seen in e.g. $o$ Andromedae \citep{Clark2003}. Whilst determining whether a double peaked emission line is a genuine shell line, rather than due to $V/R$ variability or another mechanism is non trivial, \citet{Hanuschik1996} note that \emph{only} shell-type absorption can cause the central peak to dip below the adjacent continuum, as seen in \sxp. Such an inclination would also lead to deep narrow Balmer profiles (again, as seen in \sxp ) as instead of seeing photospheric absorption profiles from the rapidly rotating star, we are in fact seeing absorption from circumstellar material at large radii (and hence low velocities).

The width of the \Halp\ profile can be used to calculate the radius of the circumstellar disc \citep{Coe2006}. Using the peak separation ( $11\pm1$~\AA{}) of the line profile as a characteristic width and assuming the circumstellar disc lies in our line of sight ($i=90^\circ$), and a stellar mass, $M_{*}$, of $18.0M_\odot$ (corresponding to a B0V star, \citealt{AstrophysicalQuantities}), the radius is given by:

\begin{equation}
\centering
R_\mathrm{circ} =\frac{GM_{*}\sin ^2 i}{(0.5\Delta v)^2}
\end{equation}

where $\Delta v$ is the peak separation in velocity space. This results in a radius of $(3.8\pm1.3)\times10^{10}$\,m  (though clearly it can be smaller if the inclination is different). We note here that the OB stars in the SMC are often overly luminous compared to Galactic systems and hence the mass could be somewhat lower and, again, the resulting disc size comparably smaller.

If the dips in $I$-band flux, discussed in Section \ref{sect:Porb}, are entirely due to some obscuring object passing in front of the circumstellar disc, then we can estimate its size using the relation:\begin{equation}
\centering
\frac{\Delta F}{F}=\frac{R^2}{AR_\mathrm{circ}^2}
\end{equation}
where $F$ is the combined flux of the star and circumstellar disc, $\Delta F$ is the drop in flux caused by the occultation, $R$ is the radius of the obscuring object, $R_\mathrm{circ}$ the radius of the circumstellar material and $A$ is a scale factor to account for the different geometries between the emitting and obscuring bodies. Rearranging for $R$ and substituting in for magnitude gives:\begin{equation}
\centering
R=AR_\mathrm{circ}\sqrt{1-10^{-0.4\Delta m}}
\end{equation} where $\Delta m$ is the change in magnitude. From Fig.~\ref{fig:ogle} we can see that $\Delta I$ varies from $\sim$0.04--0.12, which leads to a size estimate of $\sim(5-10)A\times10^{9}$~m using our estimates above for the circumstellar disc size. Determining the specific value of $A$ is not possible without more information on the nature and/or orientation of the occulting object, and the circumstellar disc but in the simplest case where both objects have the same geometry, $A=1$. One possibility is that this object is an inhomogeneity in the material in the equatorial plane (i.e. a lump), however, it is difficult to understand how a lump of material could be supported for several orbits in a differentially rotating disc. 

If we interpret the optical period as an orbital period then it follows that some feature associated with the transit of the NS is an obvious explanation for these regular features. As discussed in Section \ref{sect:intro}, the orbital periods of BeXRB are correlated with their pulse periods \citep{Corbet1984,Corbet1986}. The $40.0\pm0.3$~d is in the the expected region of an orbital period for an $\sim8$~s pulsar and would locate this source alongside other similar period sources such as SXP\,7.78 ($P_{orb}$ = 44.8~d) and SXP\,8.80 ($P_{orb}$ = 33.4~d) on the Corbet Diagram. \citet{Bird2012} present an extensive study on the periodicities present in the OGLE $I$-band light curves of the SMC BeXRBs. They show that aliasing between the non-radial pulsations of the Be star and the OGLE sampling can lead to apparent periods in the 2-100 d region, however, the strong correlation between the X-ray and optical behaviour of \sxp\ in the period MJD = 56600 -- 56660 supports the hypothesis that this period is orbital in origin.

 \citet{Rajoelimanana2011} attribute the evolution from flares and dips seen in SMC\,X-2 to a precessing elongated disc: When the semi-major axis of the Be disc extends to either side of the star (with respect to the line of sight of the observer) we see flares when the NS interacts with the disc. When the semi-major axis of the disc points towards the observer, the interaction of the NS and disc blocks light from the Be star, creating dips. Whilst there are some similarities in the behaviour of SMC\,X-2 and \sxp\ (e.g. the dips are seen as the source fades) there are some clear differences - most noticeably, that the flares are much more uniform in SMC\,X-2 (i.e. we see several cycles) and of a similar amplitude to the dips, whereas we see only a few flares in \sxp\ with $\Delta I$ ranging from 0.07--0.27.

 The value of $R$ derived above is not compatible with the radius of a NS alone (generally assumed to be 10--15~km). \citet{Maggi2013} invoke a similarly large sized obscuring object, $\ge 3.5\times10^{9}$ m, to explain the dips in the light curve of LXP\,169. They claim that these dips are due to the transit of the NS system across the Be star, rather than an obscuration of part of the circumstellar disc, and that the size can be understood in terms of the inflow of material onto the NS from the stellar wind which would create a total accreting system of the requisite size. It seems probable, whatever the detailed explanation, that \sxp\ and LXP\,169 could have much in common.

\subsection{Global Behaviour}\label{sect:discussion} 

The long term interplay between the optical and X-ray activity of \sxp\ can be easily understood within our current framework of BeXRBs. When the optical counterpart of \sxp\ is in a quiescent state there is little to no circumstellar disc present (as evidenced by the low levels of Balmer emission in the optical spectra at minimum). No material is available for accretion onto the NS and so the system is similarly dormant in X-rays. During the 1--2 year outbursts the discs grows considerably, providing a reservoir of material for the NS to accrete from. Within the most recent outburst we are seeing the same behaviour on much shorter timescales: We see rapid optical flares, occurring at $\sim P_{orb}$. From the lower panels of Fig. \ref{fig:ogle} we can see that X-ray activity closely tracks the optical during this episode, with the source in decline in the X-ray as the flare at MJD$\sim$56610 abates, falling below the detection threshold and then rising with the subsequent flare. The apparent peculiarities in the optical spectra of \sxp\ (very narrow Balmer series, deep central depression in the \Halp\ ) can be explained by the optical counterpart of \sxp\ being a shell star viewed edge on, though we note that the low resolution of the EFOSC2 spectrum means this conclusion should be treated with caution.

An open question is the cause of the flares and dips seen in the optical light curve of the 2013 activity. One possibility is that the dips are occultations of the circumstellar disc by an accretion disc that has built up around the NS. We would expect such an accretion disc to be transient, forming during accretion episodes and depleting with the circumstellar disc as it is no longer renewed. This scenario seems as odds with the \xmm\ and \swift\ non-detections: The dips in the light curve occur over a period of $\sim200$~d, suggesting that the disc is consistently present throughout this interval, however, the 2013 \xmm\ non-detection at MJD = 56597 occurs 56 days after the \cxo\ detection and suggests a drop in flux of at least two order of magnitude, followed by similar increase just 15 days later. The non detection and large dynamic range suggests that no accretion disc exists during this period. We note that several (though not all) BeXRBs in the SMC have been detected between outbursts with \lx in the range $10^{33-34}$\unitL (e.g. \citealt{Laycock2010}) implying quiescent accretion, in one form or another, can occur in a subset these systems. 

An alternate interpretation is that the flares and dips are caused by a precessing elongated circumstellar disc, like that inferred by \citet{Rajoelimanana2011} for SMC\,X-2. Again, this scenario is difficult to reconcile with the data, with the circumstellar disc appearing to disappear almost completely (as evidenced by the drop to the $I$-band base level coinciding with little to no Balmer emission in the optical spectra) mid-outburst, and re-establishing $\sim200$~d later. It is unclear how such a dramatically variable disc could support a stable super-orbital period.

\subsection{Theoretical Prospects}

Statistically, identifying a BeXRB in a shell orientation should be quite a rare occurrence. At the time of writing we are not aware of any other such system in the SMC or the LMC, though we note that we would expect LXP\,169 to be such a system if the orbital plane is perpendicular to the rotational axis of the Be star (i.e. the NS orbits in plane of circumstellar disc). To date, no optical spectra of LXP\,169 have been published.

\citet{Reig2016} present the long term spectroscopic variability of several Galactic HXMBs with data spanning the period 1987 -- 2014 (though this varies for individual sources). In their sample, the BeXRBs 4U\,0115+63/V635 Cas, RX\,J0146.9+6121/LS\,I\,+61$^\circ$\,235, RX\,J0440.9+4431/LS\,V\,+44\,17, 1A\,0535+262/V725\,Tau, IGR\,J06074+2205, GRO\,J2058+42, SAX\,J2103.5+4545, IGR\,J21343+4738 and 4U\,2206+54, have at some point, had their \Halp\ line profiles characterised as a shell profile. Their behaviours vary dramatically, with e.g. 4U\,0115+63 having shell phases of \textless~2~months whilst 4U\,2206+54 appears to be in a near constant shell configuration over the entire 24~years of monitoring - highlighting both how variable and how stable the circumstellar discs in these systems can be.

If \sxp\ is indeed in a shell orientation, this system has the potential to be a unique laboratory for understanding BeXRB systems. The very narrow line profiles of \sxp\ could be used to look for motion in the Be star itself. Combined with X-ray pulse timing analysis (see \citealt{Townsend2011}), this could be used to determine the orbital parameters of the system and hence the masses of both the components of the binary, as the inclination of the system, $i$, is already somewhat constrained. Continued OGLE monitoring of \sxp\ will allow us to identify when the disc is completely absent (e.g. MJD $\sim$ 56630 and 57030). An optical spectrum taken at such a point would allow us to determine the rotational velocity of the Be star (again, since we already have a rough idea of the inclination) which, in turn, would help to determine its pre-supernova evolutionary history. 

\section{Conclusions}

In this paper we have presented an $\sim18$ year archival X-ray and optical study of \sxp, as well as detailed study of its 2013 outburst. We revise the position of the source, based on the 2013 \cxo\ and \xmm\ data, to RA(J2000)=00:57:58.4, Dec(J2000)=-72:22:29.5 with a $1\sigma$ uncertainty of 1.5\arcsec, a position more than 20 arcminutes to the south-east to the originally reported location. This conclusively rules out the possibility that the originally identified counterpart, AzV285, is the companion star of \sxp. We identify and spectrally classify the correct counterpart; a B1Ve star with an uncharacteristically narrow Balmer series that has shown dramatic optical outbursts that coincide with the X-ray activity of \sxp. The \Halp\ emission line has a distinctive shell profile, i.e. a deep absorption core embedded in an emission line, indicating that we are likely viewing the star near edge on. An optical period of $40.0\pm0.3$~days is derived from the OGLE light curve of this star and attributed to the orbital period of the system. The X-ray spectrum is well described by an absorbed power law; typical for a BeXRB.

\section*{Acknowledgements}

We would like thank Chris Evans for making the spectra of several isolated SMC B stars available to us. ESB acknowledges support from a Claude-Leon Foundation Fellowship whilst at the University of Cape Town and from the Marie Curie Actions of the European Commission (FP7-COFUND). ESB also thanks Prof. Tom Jarrett for the generous gift of a laptop charger, which greatly facilitated work on this paper. PE acknowledges funding in the framework of the NWO Vidi award A.2320.0076.

Based on observations obtained with \xmm\, an ESA science mission with instruments and contributions directly funded by ESA Member States and NASA, and data obtained from the \emph{Chandra} Data Archive as well as observations made with ESO Telescopes at the La Silla Paranal Observatory under Program ID 092.B-0162 and observations obtained with the Southern African Large Telescope under Program 2013-2-RSA\_UKSC-001 (PI: Bartlett). The OGLE project has received funding from the National Science Centre, Poland, grant MAESTRO 2014/14/A/ST9/00121 to AU



\bibliographystyle{mnras}
\bibliography{sxp7.92} 
\bsp    

\newpage
\appendix
\section{Log of X-ray Observations included in this Work}
\begin{table*}
\caption{log of the X-ray observations of \sxp\ reported in this paper. The entries in bold denote detections. The lines mark the boundaries of the panels in Figure \ref{fig:ogle}}\label{tab:xray}
\centering
{\scriptsize \begin{tabular}{lrcclccc}
\hline\noalign{\smallskip}
Telescope 		& OBS-ID		 & MJD 		& Date 		& Instrument 		& Exp. Time 		& Flux 				& \lx$^{\dagger}$      \\\noalign{\smallskip}
		          	&			 &			&			&				& (ks)			& $\times10^{-14}$~\unitF		& \ergs{34}	\\\noalign{\smallskip}
\hline\noalign{\smallskip}

\xmm\ 			& 0110000201	 & 51834		& 2000-10-17			& EPIC-MOS2			& 5.5				& $<9.1$				& $<3.9$			\\\noalign{\smallskip}
\xmm\			& 0018540101	 & 52234		& 2001-11-21			& EPIC-MOS1			& 9.7				& $<8.8$				& $<3.8$			\\\noalign{\smallskip}
\xmm\			& 0084200101	 & 52363		& 2002-03-30			& EPIC-pn			& 4.3				& $<4.4$				& $<1.9$			\\\noalign{\smallskip}
\cxo\			& 2948		 & 52459		& 2002-07-04			& ACIS-I				& 9.4				& $<7.7$				& $<3.3$			\\\noalign{\smallskip}
\textbf{\rxte}$^{\star}$& \textbf{90086-01-01}  & \textbf{53068}		& \textbf{2004-03-04}	& \textbf{PCA}			& \textbf{12.7}		& $\mathbf{(5.9\pm2.3)\times10^3}$	& $\mathbf{(2.7\pm1.1)\times10^3}$	\\\noalign{\smallskip}
\xmm\			& 0212282601	& 53456	 	& 2005-03-27 			& EPIC-MOS1			& 0.9			& $<29.8$     & $<12.8$		\\\noalign{\smallskip}
\xmm\			& 0304250401	& 53701		& 2005-11-27			& EPIC-MOS2			& 6.5			& $<7.8$      & $<3.4$			\\\noalign{\smallskip}
\xmm\			& 0304250501	& 53703		& 2005-11-29			& EPIC-MOS2			& 6.1			& $<8.2$      & $<3.5$ 		\\\noalign{\smallskip}
\xmm\			& 0304250601    & 53715		& 2005-12-11			& EPIC-MOS1			& 5.0		        & $<9.4$      & $<4.1$			\\\noalign{\smallskip}
\swift\                        &  00030831001  &  54060        & 2006-11-21                    & XRT-pc                           &  1.4                &  $ < 1.0 \times10^2 $ &  $ < 0.4 \times10^2$ 	 \\\noalign{\smallskip}
\xmm\			& 0500980201	& 54257		& 2007-06-06			& EPIC-pn			& 10.1		& $<6.4$      & $<2.7$			\\\noalign{\smallskip}

\textbf{\rxte}$^{\star}$& \textbf{93037-01-49}	& \textbf{54630}& \textbf{2008-06-13}	& \textbf{PCA}	  & \textbf{10.2}  & $\mathbf{(1.3\pm0.5)\times10^3}$	& $\mathbf{(6.0\pm2.7)\times10^2}$		\\\noalign{\smallskip}
\textbf{\rxte}$^{\star}$& \textbf{93037-01-50}	& \textbf{54638}& \textbf{2008-06-20}	& \textbf{PCA}	  & \textbf{10.0}  & $\mathbf{(2.0\pm0.7)\times10^3}$	& $\mathbf{(9.2\pm3.2)\times10^2}$ 		\\\noalign{\smallskip}
\textbf{\rxte}$^{\star}$& \textbf{93037-01-51}	& \textbf{54646}& \textbf{2008-06-28}	& \textbf{PCA}	  & \textbf{6.6}   & $\mathbf{(1.5\pm0.5)\times10^3}$	& $\mathbf{(6.9\pm2.3)\times10^2}$ 		\\\noalign{\smallskip}
\textbf{\swift\ } 	& \textbf{00031235001}  & \textbf{54660}& \textbf{2008-07-13}   & \textbf{XRT-pc} & \textbf{1.0}   &  $\mathbf{( 2.0_{-0.5}^{+0.6})\times10^2}$  &  $\mathbf{( 0.9_{-0.2}^{+0.3} )\times10^2}$ 	 \\\noalign{\smallskip}

\swift\ &  00031237001  &  54660  &  2008-07-13  & XRT-pc    &  1.2   &  $ < 1.9 \times10^2 $ &  $ < 0.8 \times10^2$ 	 \\\noalign{\smallskip}
\swift\ &  00035413001  &  54782  &  2008-11-12  & XRT-pc    &  11.8  &  $ < 0.2 \times10^2 $ &  $ < 0.1 \times10^2$ 	 \\\noalign{\smallskip}
\xmm\ &  0601210801 &  55113  &  2009-10-09  & EPIC-pn   & 12.5   &  $ < 2.7            $ &  $ < 1.2$        \\\noalign{\smallskip}
\swift\ &  00040419001  &  55467  &  2010-09-28  & XRT-pc    &  3.2   &  $ < 34             $ &  $ < 15 $ 	 \\\noalign{\smallskip}
\swift\ &  00040430003  &  55894  &  2011-11-29  & XRT-pc    &  1.0   &  $ < 2.4 \times10^2 $ &  $ < 1.0 \times10^2$ 	 \\\noalign{\smallskip}

\hline\noalign{\smallskip}

\textbf{\cxo\ }		& \textbf{14671}	& \textbf{56299}	& \textbf{2013-01-07}	& \textbf{ACIS-I}	& \textbf{48.8}	& $\mathbf{32\pm2}$		        & $\mathbf{14.7\pm0.9}$		\\\noalign{\smallskip}
\textbf{\cxo\ }		& \textbf{13773}	& \textbf{56332}	& \textbf{2013-02-09}	& \textbf{ACIS-S}	& \textbf{39.1}	& $\mathbf{29\pm2}$	       	        & $\mathbf{13.3\pm0.9}$		\\\noalign{\smallskip}
\textbf{\cxo\ }		& \textbf{15504}	& \textbf{56541}	& \textbf{2013-09-06}	& \textbf{ACIS-I}	& \textbf{49.5}	& $\mathbf{(2.05\pm0.07)\times10^2}$	& $\mathbf{94.3\pm3.2}$		\\\noalign{\smallskip}
\xmm\			& 0700580101	   	& 56597			& 2013-11-01			& EPIC-MOS2     & 13.3			& $<3.2$				      & $<1.4$		\\\noalign{\smallskip}

\hline\noalign{\smallskip}

\swift\                        &  00040419002  	&  56601 		        &  2013-11-05                 	& XRT-pc        	   &  0.8                      &  $ < 1.4 \times10^2 $           &  $ < 0.6 \times10^2$ \\\noalign{\smallskip}
\textbf{\swift\ } & \textbf{00033038001}  &  \textbf{56612}  &  \textbf{2013-11-16}  & \textbf{XRT-pc}    &  \textbf{1.0}  &  $\mathbf{( 6.8\pm1.0 )\times10^2}$  &  $\mathbf{( 2.9\pm0.4 )\times10^2}$ \\\noalign{\smallskip}
\textbf{\swift\ } & \textbf{00033038002}  &  \textbf{56614}  &  \textbf{2013-11-18}  & \textbf{XRT-pc}    &  \textbf{1.0}  &  $\mathbf{( 4.3\pm0.8 )\times10^2}$  &  $\mathbf{( 1.8\pm0.4 )\times10^2}$ \\\noalign{\smallskip}
\textbf{\xmm\ } & \textbf{0700580401}  &  \textbf{56616}  &  \textbf{2013-11-20}  & \textbf{EPIC-pn}	&  \textbf{16.7}  & $\mathbf{(2.80^{+0.07}_{-0.10})\times10^2}$& $\mathbf{(1.29^{+0.03}_{-0.04})\times10^2}$ \\\noalign{\smallskip}
\textbf{\swift\ } & \textbf{00033038003}  &  \textbf{56616}  &  \textbf{2013-11-20}  & \textbf{XRT-pc}    &  \textbf{0.4}  &  $\mathbf{( 6.4\pm1.6 )\times10^2}$  &  $\mathbf{( 2.8\pm0.7 )\times10^2}$ \\\noalign{\smallskip}
\textbf{\swift\ } & \textbf{00033038004}  &  \textbf{56618}  &  \textbf{2013-11-22}  & \textbf{XRT-pc}    &  \textbf{1.1}  &  $\mathbf{( 2.3\pm0.6 )\times10^2}$  &  $\mathbf{( 1.0\pm0.3 )\times10^2}$ \\\noalign{\smallskip}
\textbf{\swift\ } & \textbf{00033038005}  &  \textbf{56620}  &  \textbf{2013-11-24}  & \textbf{XRT-pc}    &  \textbf{0.9}  &  $\mathbf{( 3.2\pm0.7 )\times10^2}$  &  $\mathbf{( 1.4\pm0.3 )\times10^2}$ \\\noalign{\smallskip}
\textbf{\swift\ } & \textbf{00033038006}  &  \textbf{56622}  &  \textbf{2013-11-26}  & \textbf{XRT-pc}    &  \textbf{0.9}  &  $\mathbf{( 1.6_{-0.5}^{+0.6})\times10^2}$  &  $\mathbf{( 0.7_{-0.2}^{+0.3} )\times10^2}$ \\\noalign{\smallskip}

\swift\ &  00033038008  &  56626  &  2013-11-30  & XRT-pc &  1.2   &  $ < 1.7 \times10^2 $ &  $ < 0.7 \times10^2$ 	 \\\noalign{\smallskip}
\swift\ &  00033038009  &  56628  &  2013-12-02  & XRT-pc &  1.0   &  $ < 1.6 \times10^2 $ &  $ < 0.7 \times10^2$ 	 \\\noalign{\smallskip}
\swift\ &  00033038010  &  56630  &  2013-12-04  & XRT-pc &  0.8   &  $ < 2.0 \times10^2 $ &  $ < 0.9 \times10^2$ 	 \\\noalign{\smallskip}
\swift\ &  00033038011  &  56630  &  2013-12-04  & XRT-pc &  1.0   &  $ < 1.0 \times10^2 $ &  $ < 0.4 \times10^2$ 	 \\\noalign{\smallskip}
\swift\ &  00033038012  &  56631  &  2013-12-05  & XRT-pc &  0.9   &  $ < 1.2 \times10^2 $ &  $ < 0.5 \times10^2$ 	 \\\noalign{\smallskip}
\swift\ &  00033038013  &  56631  &  2013-12-05  & XRT-pc &  0.9   &  $ < 1.5 \times10^2 $ &  $ < 0.7 \times10^2$ 	 \\\noalign{\smallskip}
\swift\ &  00033038015  &  56632  &  2013-12-06  & XRT-pc &  1.0   &  $ < 1.0 \times10^2 $ &  $ < 0.4 \times10^2$ 	 \\\noalign{\smallskip}
\swift\ &  00033038014  &  56632  &  2013-12-06  & XRT-pc &  1.0   &  $ < 1.3 \times10^2 $ &  $ < 0.5 \times10^2$ 	 \\\noalign{\smallskip}
\swift\ &  00033038016  &  56633  &  2013-12-07  & XRT-pc &  1.2   &  $ < 1.1 \times10^2 $ &  $ < 0.5 \times10^2$ 	 \\\noalign{\smallskip}
\swift\ &  00033038017  &  56635  &  2013-12-09  & XRT-pc &  1.0   &  $ < 1.2 \times10^2 $ &  $ < 0.5 \times10^2$ 	 \\\noalign{\smallskip}
\swift\ &  00033038018  &  56637  &  2013-12-11  & XRT-pc &  0.9   &  $ < 1.5 \times10^2 $ &  $ < 0.6 \times10^2$ 	 \\\noalign{\smallskip}
\textbf{\swift\ } & \textbf{00033038020}  &  \textbf{56642}  &  \textbf{2013-12-16}  & \textbf{XRT-pc}    &  \textbf{1.2}  &  $\mathbf{( 1.6_{-0.4}^{+0.5})\times10^2}$  &  $\mathbf{( 0.7_{-0.2}^{+0.2} )\times10^2}$\\\noalign{\smallskip}
\swift\ &  00033038022  &  56643  &  2013-12-17  & XRT-pc &  0.9   &  $ < 1.0 \times10^2 $ &  $ < 0.4 \times10^2$ 	 \\\noalign{\smallskip}
\swift\ &  00033038023  &  56643  &  2013-12-17  & XRT-pc &  1.0   &  $ < 1.4 \times10^2 $ &  $ < 0.6 \times10^2$ 	 \\\noalign{\smallskip}
\swift\ &  00033038024  &  56643  &  2013-12-17  & XRT-pc &  0.9   &  $ < 2.3 \times10^2 $ &  $ < 1.0 \times10^2$ 	 \\\noalign{\smallskip}
\swift\ &  00033038025  &  56643  &  2013-12-17  & XRT-pc &  1.0   &  $ < 1.7 \times10^2 $ &  $ < 0.7 \times10^2$ 	 \\\noalign{\smallskip}
\swift\ &  00033038027  &  56643  &  2013-12-17  & XRT-pc &  0.9   &  $ < 1.5 \times10^2 $ &  $ < 0.6 \times10^2$ 	 \\\noalign{\smallskip}
\swift\ &  00033038028  &  56644  &  2013-12-18  & XRT-pc &  0.9   &  $ < 1.0 \times10^2 $ &  $ < 0.4 \times10^2$ 	 \\\noalign{\smallskip}
\swift\ &  00033038029  &  56644  &  2013-12-18  & XRT-pc &  0.8   &  $ < 1.3 \times10^2 $ &  $ < 0.6 \times10^2$ 	 \\\noalign{\smallskip}
\swift\ &  00033038030  &  56644  &  2013-12-18  & XRT-pc &  1.0   &  $ < 1.0 \times10^2 $ &  $ < 0.4 \times10^2$ 	 \\\noalign{\smallskip}
\swift\ &  00033038031  &  56644  &  2013-12-18  & XRT-pc &  0.7   &  $ < 1.9 \times10^2 $ &  $ < 0.8 \times10^2$ 	 \\\noalign{\smallskip}
\swift\ &  00033038032  &  56644  &  2013-12-18  & XRT-pc &  1.0   &  $ < 1.0 \times10^2 $ &  $ < 0.4 \times10^2$ 	 \\\noalign{\smallskip}
\swift\ &  00033038033  &  56644  &  2013-12-18  & XRT-pc &  1.0   &  $ < 1.6 \times10^2 $ &  $ < 0.7 \times10^2$ 	 \\\noalign{\smallskip}

\hline
\end{tabular}}
\end{table*}

\begin{table*}
{\scriptsize\centering
\begin{tabular}{lrcclccc}
\hline\noalign{\smallskip}
Telescope 		& OBS-ID		 & MJD 		& Date 		& Instrument 		& Exp. Time 		& Flux 				& \lx$^{\dagger}$      \\\noalign{\smallskip}
		      	 	&			 &			&			&				& (ks)			& $\times10^{-14}$~\unitF		& \ergs{34}	\\\noalign{\smallskip}
\hline\noalign{\smallskip}
\swift\ &  00033038034  &  56645  &  2013-12-19  & XRT-pc &  1.0   &  $ < 1.0 \times10^2 $ &  $ < 0.4 \times10^2$ 	 \\\noalign{\smallskip}
\swift\ &  00033038035  &  56645  &  2013-12-19  & XRT-pc &  0.9   &  $ < 1.8 \times10^2 $ &  $ < 0.8 \times10^2$ 	 \\\noalign{\smallskip}
\swift\ &  00033038038  &  56645  &  2013-12-19  & XRT-pc &  0.7   &  $ < 2.5 \times10^2 $ &  $ < 1.1 \times10^2$ 	 \\\noalign{\smallskip}

\textbf{\swift\ } & \textbf{00033038036}  &  \textbf{56645}  &  \textbf{2013-12-19}  & \textbf{XRT-pc}    &  \textbf{1.1}  &  $\mathbf{( 0.6_{-0.3}^{+0.4})\times10^2}$  &  $\mathbf{( 0.3_{-0.1}^{+0.2} )\times10^2}$\\\noalign{\smallskip}
\textbf{\swift\ } & \textbf{00033038037}  &  \textbf{56645}  &  \textbf{2013-12-19}  & \textbf{XRT-pc}    &  \textbf{1.0}  &  $\mathbf{( 0.8_{-0.3}^{+0.4})\times10^2}$  &  $\mathbf{( 0.3_{-0.1}^{+0.2} )\times10^2}$\\\noalign{\smallskip}
\textbf{\swift\ } & \textbf{00033038039}  &  \textbf{56645}  &  \textbf{2013-12-19}  & \textbf{XRT-pc}    &  \textbf{0.9}  &  $\mathbf{( 1.0_{-0.5}^{+0.6})\times10^2}$  &  $\mathbf{( 0.4_{-0.2}^{+0.3} )\times10^2}$\\\noalign{\smallskip}
\textbf{\swift\ } & \textbf{00033038040}  &  \textbf{56646}  &  \textbf{2013-12-20}  & \textbf{XRT-pc}    &  \textbf{1.0}  &  $\mathbf{( 3.3\pm0.7 )\times10^2}$  &  $\mathbf{( 1.4\pm0.3 )\times10^2}$ 	 \\\noalign{\smallskip}
\textbf{\swift\ } & \textbf{00033038041}  &  \textbf{56646}  &  \textbf{2013-12-20}  & \textbf{XRT-pc}    &  \textbf{0.9}  &  $\mathbf{( 2.0_{-0.5}^{+0.7})\times10^2}$  &  $\mathbf{( 0.9_{-0.2}^{+0.3} )\times10^2}$\\\noalign{\smallskip}
\textbf{\swift\ } & \textbf{00033038042}  &  \textbf{56646}  &  \textbf{2013-12-20}  & \textbf{XRT-pc}    &  \textbf{1.0}  &  $\mathbf{( 1.2_{-0.4}^{+0.5})\times10^2}$  &  $\mathbf{( 0.5_{-0.2}^{+0.2} )\times10^2}$\\\noalign{\smallskip}
\textbf{\swift\ } & \textbf{00033038043}  &  \textbf{56646}  &  \textbf{2013-12-20}  & \textbf{XRT-pc}    &  \textbf{1.0}  &  $\mathbf{( 2.3\pm0.6 )\times10^2}$  &  $\mathbf{( 1.0\pm0.3 )\times10^2}$\\\noalign{\smallskip}
\textbf{\swift\ } & \textbf{00033038044}  &  \textbf{56646}  &  \textbf{2013-12-20}  & \textbf{XRT-pc}    &  \textbf{0.7}  &  $\mathbf{( 3.0_{-0.9}^{+1.2})\times10^2}$  &  $\mathbf{( 1.3_{-0.4}^{+0.5} )\times10^2}$\\\noalign{\smallskip}
\textbf{\swift\ } & \textbf{00033038045}  &  \textbf{56646}  &  \textbf{2013-12-20}  & \textbf{XRT-pc}    &  \textbf{1.0}  &  $\mathbf{( 2.5\pm0.6 )\times10^2}$  &  $\mathbf{( 1.1\pm0.3 )\times10^2}$ 	 \\\noalign{\smallskip}
\textbf{\swift\ } & \textbf{00033038046}  &  \textbf{56647}  &  \textbf{2013-12-21}  & \textbf{XRT-pc}    &  \textbf{1.0}  &  $\mathbf{( 2.1_{-0.6}^{+0.7})\times10^2}$  &  $\mathbf{( 0.9_{-0.3}^{+0.3} )\times10^2}$\\\noalign{\smallskip}
\textbf{\swift\ } & \textbf{00033038047}  &  \textbf{56647}  &  \textbf{2013-12-21}  & \textbf{XRT-pc}    &  \textbf{0.8}  &  $\mathbf{( 2.5_{-0.6}^{+0.8})\times10^2}$  &  $\mathbf{( 1.1_{-0.3}^{+0.3} )\times10^2}$\\\noalign{\smallskip}
\textbf{\swift\ } & \textbf{00033038048}  &  \textbf{56647}  &  \textbf{2013-12-21}  & \textbf{XRT-pc}    &  \textbf{1.0}  &  $\mathbf{( 4.2\pm0.8 )\times10^2}$  &  $\mathbf{( 1.8\pm0.4 )\times10^2}$\\\noalign{\smallskip}
\textbf{\swift\ } & \textbf{00033038049}  &  \textbf{56647}  &  \textbf{2013-12-21}  & \textbf{XRT-pc}    &  \textbf{1.0}  &  $\mathbf{( 4.1_{-1.1}^{+1.3})\times10^2}$  &  $\mathbf{( 1.8_{-0.5}^{+0.6} )\times10^2}$\\\noalign{\smallskip}
\textbf{\swift\ } & \textbf{00033038050}  &  \textbf{56647}  &  \textbf{2013-12-21}  & \textbf{XRT-pc}    &  \textbf{1.0}  &  $\mathbf{( 5.4\pm1.0 )\times10^2}$  &  $\mathbf{( 2.3\pm0.4 )\times10^2}$ 	 \\\noalign{\smallskip}
\textbf{\swift\ } & \textbf{00033038051}  &  \textbf{56647}  &  \textbf{2013-12-21}  & \textbf{XRT-pc}    &  \textbf{1.0}  &  $\mathbf{( 3.3\pm0.7 )\times10^2}$  &  $\mathbf{( 1.4\pm0.3 )\times10^2}$ 	 \\\noalign{\smallskip}
\textbf{\swift\ } & \textbf{00033038052}  &  \textbf{56648}  &  \textbf{2013-12-22}  & \textbf{XRT-pc}    &  \textbf{0.9}  &  $\mathbf{( 6.2\pm1.0 )\times10^2}$  &  $\mathbf{( 2.6\pm0.4 )\times10^2}$ 	 \\\noalign{\smallskip}
\textbf{\swift\ } & \textbf{00033038053}  &  \textbf{56648}  &  \textbf{2013-12-22}  & \textbf{XRT-pc}    &  \textbf{0.6}  &  $\mathbf{( 1.7_{-0.6}^{+0.8})\times10^2}$  &  $\mathbf{( 0.7_{-0.3}^{+0.3} )\times10^2}$\\\noalign{\smallskip}
\textbf{\swift\ } & \textbf{00033038054}  &  \textbf{56648}  &  \textbf{2013-12-22}  & \textbf{XRT-pc}    &  \textbf{1.0}  &  $\mathbf{( 4.5\pm0.8 )\times10^2}$  &  $\mathbf{( 2.0\pm0.4 )\times10^2}$ 	 \\\noalign{\smallskip}
\textbf{\swift\ } & \textbf{00033038056}  &  \textbf{56648}  &  \textbf{2013-12-22}  & \textbf{XRT-pc}    &  \textbf{0.9}  &  $\mathbf{( 4.0\pm0.8 )\times10^2}$  &  $\mathbf{( 1.7\pm0.3 )\times10^2}$ 	 \\\noalign{\smallskip}
\textbf{\swift\ } & \textbf{00033038057}  &  \textbf{56648}  &  \textbf{2013-12-22}  & \textbf{XRT-pc}    &  \textbf{1.0}  &  $\mathbf{( 2.5\pm0.6 )\times10^2}$  &  $\mathbf{( 1.1\pm0.3 )\times10^2}$ 	 \\\noalign{\smallskip}
\textbf{\swift\ } & \textbf{00033038058}  &  \textbf{56649}  &  \textbf{2013-12-23}  & \textbf{XRT-pc}    &  \textbf{0.8}  &  $\mathbf{( 3.2\pm0.8 )\times10^2}$  &  $\mathbf{( 1.4\pm0.3 )\times10^2}$ 	 \\\noalign{\smallskip}
\textbf{\swift\ } & \textbf{00033038059}  &  \textbf{56649}  &  \textbf{2013-12-23}  & \textbf{XRT-pc}    &  \textbf{0.9}  &  $\mathbf{( 3.8_{-1.1}^{+1.3})\times10^2}$  &  $\mathbf{( 1.6_{-0.5}^{+0.6} )\times10^2}$\\\noalign{\smallskip}
\textbf{\swift\ } & \textbf{00033038060}  &  \textbf{56649}  &  \textbf{2013-12-23}  & \textbf{XRT-pc}    &  \textbf{1.0}  &  $\mathbf{( 4.0\pm0.8 )\times10^2}$  &  $\mathbf{( 1.7\pm0.4 )\times10^2}$ 	 \\\noalign{\smallskip}
\textbf{\swift\ } & \textbf{00033038061}  &  \textbf{56649}  &  \textbf{2013-12-23}  & \textbf{XRT-pc}    &  \textbf{0.8}  &  $\mathbf{( 3.8_{-1.1}^{+1.3})\times10^2}$  &  $\mathbf{( 1.6_{-0.5}^{+0.6} )\times10^2}$\\\noalign{\smallskip}
\textbf{\swift\ } & \textbf{00033038062}  &  \textbf{56649}  &  \textbf{2013-12-23}  & \textbf{XRT-pc}    &  \textbf{0.7}  &  $\mathbf{( 4.5\pm1.0 )\times10^2}$  &  $\mathbf{( 1.9\pm0.4 )\times10^2}$ 	 \\\noalign{\smallskip}
\textbf{\swift\ } & \textbf{00033038063}  &  \textbf{56649}  &  \textbf{2013-12-23}  & \textbf{XRT-pc}    &  \textbf{0.9}  &  $\mathbf{( 4.3\pm0.9 )\times10^2}$  &  $\mathbf{( 1.9\pm0.4 )\times10^2}$ 	 \\\noalign{\smallskip}
\textbf{\swift\ } & \textbf{00033038064}  &  \textbf{56650}  &  \textbf{2013-12-24}  & \textbf{XRT-pc}    &  \textbf{1.0}  &  $\mathbf{( 3.9\pm0.8 )\times10^2}$  &  $\mathbf{( 1.7\pm0.4 )\times10^2}$ 	 \\\noalign{\smallskip}
\textbf{\swift\ } & \textbf{00033038065}  &  \textbf{56650}  &  \textbf{2013-12-24}  & \textbf{XRT-pc}    &  \textbf{0.6}  &  $\mathbf{( 6.3\pm1.4 )\times10^2}$  &  $\mathbf{( 2.7\pm0.6 )\times10^2}$ 	 \\\noalign{\smallskip}
\textbf{\swift\ } & \textbf{00033038066}  &  \textbf{56650}  &  \textbf{2013-12-24}  & \textbf{XRT-pc}    &  \textbf{1.0}  &  $\mathbf{( 5.0\pm0.9 )\times10^2}$  &  $\mathbf{( 2.1\pm0.4 )\times10^2}$ 	 \\\noalign{\smallskip}
\textbf{\swift\ } & \textbf{00033038067}  &  \textbf{56650}  &  \textbf{2013-12-24}  & \textbf{XRT-pc}    &  \textbf{1.0}  &  $\mathbf{( 5.3\pm1.0 )\times10^2}$  &  $\mathbf{( 2.3\pm0.4 )\times10^2}$ 	 \\\noalign{\smallskip}
\textbf{\swift\ } & \textbf{00033038068}  &  \textbf{56650}  &  \textbf{2013-12-24}  & \textbf{XRT-pc}    &  \textbf{0.6}  &  $\mathbf{( 5.7\pm1.2 )\times10^2}$  &  $\mathbf{( 2.4\pm0.5 )\times10^2}$ 	 \\\noalign{\smallskip}
\textbf{\swift\ } & \textbf{00033038070}  &  \textbf{56651}  &  \textbf{2013-12-25}  & \textbf{XRT-pc}    &  \textbf{0.8}  &  $\mathbf{( 6.4\pm1.1 )\times10^2}$  &  $\mathbf{( 2.8\pm0.5 )\times10^2}$ 	 \\\noalign{\smallskip}
\textbf{\swift\ } & \textbf{00033038071}  &  \textbf{56651}  &  \textbf{2013-12-25}  & \textbf{XRT-pc}    &  \textbf{0.8}  &  $\mathbf{( 5.1\pm1.0 )\times10^2}$  &  $\mathbf{( 2.2\pm0.4 )\times10^2}$ 	 \\\noalign{\smallskip}
\textbf{\swift\ } & \textbf{00033038072}  &  \textbf{56651}  &  \textbf{2013-12-25}  & \textbf{XRT-pc}    &  \textbf{0.9}  &  $\mathbf{( 5.3\pm1.0 )\times10^2}$  &  $\mathbf{( 2.3\pm0.4 )\times10^2}$ 	 \\\noalign{\smallskip}
\textbf{\swift\ } & \textbf{00033038073}  &  \textbf{56651}  &  \textbf{2013-12-25}  & \textbf{XRT-pc}    &  \textbf{0.9}  &  $\mathbf{( 7.4\pm1.1 )\times10^2}$  &  $\mathbf{( 3.2\pm0.5 )\times10^2}$ 	 \\\noalign{\smallskip}
\textbf{\swift\ } & \textbf{00033038074}  &  \textbf{56651}  &  \textbf{2013-12-25}  & \textbf{XRT-pc}    &  \textbf{0.5}  &  $\mathbf{( 7.8\pm2.0 )\times10^2}$  &  $\mathbf{( 3.4\pm0.9 )\times10^2}$ 	 \\\noalign{\smallskip}
\textbf{\swift\ } & \textbf{00033038075}  &  \textbf{56651}  &  \textbf{2013-12-25}  & \textbf{XRT-pc}    &  \textbf{0.8}  &  $\mathbf{( 5.4\pm1.1 )\times10^2}$  &  $\mathbf{( 2.3\pm0.5 )\times10^2}$ 	 \\\noalign{\smallskip}

\textbf{\swift\ } & \textbf{00033038076}  &  \textbf{56652}  &  \textbf{2013-12-26}  & \textbf{XRT-pc}    &  \textbf{0.8}  &  $\mathbf{( 6.4\pm1.2 )\times10^2}$  &  $\mathbf{( 2.7\pm0.5 )\times10^2}$ 	 \\\noalign{\smallskip}
\textbf{\swift\ } & \textbf{00033038077}  &  \textbf{56652}  &  \textbf{2013-12-26}  & \textbf{XRT-pc}    &  \textbf{1.1}  &  $\mathbf{( 7.4\pm1.0 )\times10^2}$  &  $\mathbf{( 3.2\pm0.4 )\times10^2}$ 	 \\\noalign{\smallskip}
\textbf{\swift\ } & \textbf{00033038078}  &  \textbf{56652}  &  \textbf{2013-12-26}  & \textbf{XRT-pc}    &  \textbf{1.0}  &  $\mathbf{( 7.4\pm1.1 )\times10^2}$  &  $\mathbf{( 3.2\pm0.5 )\times10^2}$ 	 \\\noalign{\smallskip}
\textbf{\swift\ } & \textbf{00033038079}  &  \textbf{56652}  &  \textbf{2013-12-26}  & \textbf{XRT-pc}    &  \textbf{0.9}  &  $\mathbf{( 6.6\pm1.1 )\times10^2}$  &  $\mathbf{( 2.8\pm0.5 )\times10^2}$ 	 \\\noalign{\smallskip}
\textbf{\swift\ } & \textbf{00033038080}  &  \textbf{56652}  &  \textbf{2013-12-26}  & \textbf{XRT-pc}    &  \textbf{0.9}  &  $\mathbf{( 6.3\pm1.0 )\times10^2}$  &  $\mathbf{( 2.7\pm0.4 )\times10^2}$ 	 \\\noalign{\smallskip}
\textbf{\swift\ } & \textbf{00033038081}  &  \textbf{56652}  &  \textbf{2013-12-26}  & \textbf{XRT-pc}    &  \textbf{0.8}  &  $\mathbf{( 5.9\pm1.1 )\times10^2}$  &  $\mathbf{( 2.5\pm0.5 )\times10^2}$ 	 \\\noalign{\smallskip}
\textbf{\xmm\ } & \textbf{0700580601}     &  \textbf{56652}  &  \textbf{2013-12-26}  & \textbf{EPIC-MOS2} &  \textbf{18.5}   & $\mathbf{(4.10^{+0.08}_{-0.10})\times10^2}$ & $\mathbf{(1.89^{+0.03}_{-0.04})\times10^2}$ \\\noalign{\smallskip}
\textbf{\swift\ } & \textbf{00033038082}  &  \textbf{56653}  &  \textbf{2013-12-27}  & \textbf{XRT-pc}    &  \textbf{0.6}  &  $\mathbf{( 7.1\pm1.4 )\times10^2}$  &  $\mathbf{( 3.0\pm0.6 )\times10^2}$ 	 \\\noalign{\smallskip}
\textbf{\swift\ } & \textbf{00033038083}  &  \textbf{56653}  &  \textbf{2013-12-27}  & \textbf{XRT-pc}    &  \textbf{0.5}  &  $\mathbf{( 6.8\pm1.5 )\times10^2}$  &  $\mathbf{( 2.9\pm0.6 )\times10^2}$ 	 \\\noalign{\smallskip}
\textbf{\swift\ } & \textbf{00033038084}  &  \textbf{56653}  &  \textbf{2013-12-27}  & \textbf{XRT-pc}    &  \textbf{1.1}  &  $\mathbf{( 5.7\pm0.9 )\times10^2}$  &  $\mathbf{( 2.5\pm0.4 )\times10^2}$ 	 \\\noalign{\smallskip}
\textbf{\swift\ } & \textbf{00033038085}  &  \textbf{56653}  &  \textbf{2013-12-27}  & \textbf{XRT-pc}    &  \textbf{0.6}  &  $\mathbf{( 3.6\pm1.0 )\times10^2}$  &  $\mathbf{( 1.5\pm0.4 )\times10^2}$ 	 \\\noalign{\smallskip}

\swift\ &  00033038086  &  56666  &  2014-01-09  & XRT-pc &  0.8   &  $ < 2.3 \times10^2 $ &  $ < 1.0 \times10^2$ 	 \\\noalign{\smallskip}
\swift\ &  00033038087  &  56668  &  2014-01-11  & XRT-pc &  1.0   &  $ < 1.1 \times10^2 $ &  $ < 0.5 \times10^2$ 	 \\\noalign{\smallskip}
\swift\ &  00033038088  &  56670  &  2014-01-13  & XRT-pc &  1.2   &  $ < 1.5 \times10^2 $ &  $ < 0.7 \times10^2$ 	 \\\noalign{\smallskip}
\swift\ &  00033038098  &  56689  &  2014-02-01  & XRT-pc &  0.1   &  $ < 16.4 \times10^2 $ &  $ < 7.1 \times10^2$ \\\noalign{\smallskip}
\swift\ &  00033038099  &  56691  &  2014-02-03  & XRT-pc &  0.1   &  $ < 20.2 \times10^2 $ &  $ < 8.7 \times10^2$ \\\noalign{\smallskip}
\swift\ &  00033038104  &  56698  &  2014-02-10  & XRT-pc &  0.1   &  $ < 20.6 \times10^2 $ &  $ < 8.9 \times10^2$ \\\noalign{\smallskip}
\swift\ &  00033171001  &  56716  &  2014-02-28  & XRT-pc &  0.2   &  $ < 6.1 \times10^2 $ &  $ < 2.6 \times10^2$ 	\\\noalign{\smallskip}
\swift\ &  00033038108  &  56716  &  2014-02-28  & XRT-pc &  1.0   &  $ < 1.8 \times10^2 $ &  $ < 0.8 \times10^2$ 	\\\noalign{\smallskip}
\hline
\end{tabular}}
\end{table*}

\begin{table*}
{\scriptsize\centering
\begin{tabular}{lrcclccc}
\hline\noalign{\smallskip}
Telescope 		& OBS-ID		 & MJD 		& Date 		& Instrument 		& Exp. Time 		& Flux 				& \lx$^{\dagger}$      \\\noalign{\smallskip}
			       	&			 &			&			&				& (ks)			& $\times10^{-14}$~\unitF		& \ergs{34}	\\\noalign{\smallskip}
\hline\noalign{\smallskip}
\textbf{\swift\ } & \textbf{00033038110}  &  \textbf{56721}  &  \textbf{2014-03-05}  & \textbf{XRT-pc}    &  \textbf{0.6}  &  $\mathbf{( 1.7_{-0.6}^{+0.8})\times10^2}$  &  $\mathbf{( 0.7_{-0.3}^{+0.3} )\times10^2}$\\\noalign{\smallskip}

\swift\ &  00033038111  &  56723  &  2014-03-07  & XRT-pc &  1.0   &  $ < 1.8 \times10^2 $ &  $ < 0.8 \times10^2$ 	 \\\noalign{\smallskip}

\textbf{\swift\ } & \textbf{00033038112}  &  \textbf{56725}  &  \textbf{2014-03-09}  & \textbf{XRT-pc}    &  \textbf{1.3}  &  $\mathbf{( 1.3_{-0.4}^{+0.5})\times10^2}$  &  $\mathbf{( 0.6_{-0.2}^{+0.2} )\times10^2}$\\\noalign{\smallskip}
\textbf{\swift\ } & \textbf{00033038113}  &  \textbf{56727}  &  \textbf{2014-03-11}  & \textbf{XRT-pc}    &  \textbf{1.2}  &  $\mathbf{( 2.6\pm0.6 )\times10^2}$  &  $\mathbf{( 1.1\pm0.3 )\times10^2}$ 	 \\\noalign{\smallskip}

\hline\noalign{\smallskip}
\swift\ &  00033747001  &  57139  &  2015-04-27  & XRT-pc &  1.1   &  $ < 1.7 \times10^2 $ &  $ < 0.7 \times10^2$ 	 \\\noalign{\smallskip}
\hline
\end{tabular}}
\flushleft{\textbf{Notes.} ($^{\dagger}$) Assuming a distance to the SMC of 62.1 kpc \citep{Graczyk2014}. ($^{\star}$) Due to the lack of imaging capabilities of \rxte, it is impossible to accurately determine the background level and hence the absolute flux value of the source. These flux values are based on an assumed pulsed fraction of $0.3\pm0.1$.}
\end{table*}

\label{lastpage}
\end{document}